\documentclass[12pt,a4paper]{article}
\usepackage[a4paper,margin=1in]{geometry}
\usepackage{times}
\usepackage{setspace}
\usepackage{hyperref}
\usepackage{titlesec}
\usepackage{graphicx}
\usepackage{amsmath, amssymb}
\usepackage{booktabs} 

\titleformat{\section}{\large\bfseries}{\thesection.}{0.5em}{}
\titleformat{\subsection}{\normalsize\bfseries}{\thesubsection.}{0.5em}{}

\title{\textbf{From Heisenberg and Schrödinger to the \texorpdfstring{$\mathsf{P}$ vs.\ $\mathsf{NP}$}{P vs NP} Problem}}

\medskip

\author{Galina Weinstein\thanks{The Department of Philosophy, University of Haifa.}}
\date{}

\begin{document}
\maketitle
\doublespacing

\begin{abstract}
This essay offers an epistemological reinterpretation of the foundational divide between matrix mechanics and wave mechanics. 
Though formally equivalent, the two theories embody distinct modes of knowing: \emph{procedural construction} and \emph{recognitional verification}. 
These epistemic architectures anticipate, in philosophical form, the logical asymmetry expressed by the $\mathsf{P}$ versus $\mathsf{NP}$ problem in computational complexity. 
Here, the contrast between efficient generation and efficient recognition is treated not as a mathematical taxonomy but as a framework for understanding how knowledge is produced and validated across physics, computation, and cognition. 
The essay reconstructs the mathematical history of quantum mechanics through the original derivations of Werner Heisenberg, Max Born, Pascual Jordan, Paul Dirac, Erwin Schrödinger, Paul Ehrenfest, and Wolfgang Pauli, culminating in John von Neumann’s unification of both approaches within the formalism of Hilbert space. 
By juxtaposing Heisenberg’s algorithmic formalism with Schrödinger’s representational one, it argues that their divergence reveals a structural feature of scientific reasoning itself—the enduring tension between what can be procedurally constructed and what can only be recognized.
\end{abstract}

\newpage

\tableofcontents

\section{Introduction}

Computational complexity theory, once a technical branch of theoretical computer science, has become increasingly relevant to epistemology and the philosophy of science \cite{Aaronson2025}. 
Its central open question—whether $\mathsf{P}$ equals $\mathsf{NP}$—asks whether every truth that can be \emph{verified} efficiently can also be \emph{constructed}. 
Recast philosophically, it asks whether recognition and generation, understanding and production, can ever coincide.

Throughout this essay, $\mathsf{P}$ and $\mathsf{NP}$ are used not as literal complexity classes but as epistemic metaphors. 
They designate two recurrent modes of knowing across physics, computation, and cognition: \emph{procedural construction} and \emph{recognitional verification}. 
Complexity theory thus offers a precise vocabulary for an enduring philosophical dichotomy. 
No claim is made that physical theories or cognitive processes can be formally classified within these classes; rather, the distinction serves as a conceptual framework for articulating how knowledge is generated and recognized.

A similar epistemic tension shaped the birth of quantum mechanics in the 1920s. 
Werner Heisenberg’s matrix mechanics and Erwin Schrödinger’s wave mechanics provided empirically equivalent yet conceptually divergent accounts of atomic structure. 
Schrödinger himself emphasized this contrast in the opening of his 1926 paper on the relation between the two theories. 
He observed that it was “very strange” that two formalisms so different in their “starting-points, presentations, and methods”—one replacing continuous variables with discrete matrices, the other turning point-mechanics into a continuum theory—could lead to the same empirical results \cite{Schrod5}. 

This essay offers a new kind of reading: rather than treating that contrast as one between continuity and discreteness, it interprets their divergence as the emergence of two distinct epistemological modes—\emph{procedural} and \emph{recognitional}—that prefigure the logical structure of the $\mathsf{P}$ versus $\mathsf{NP}$ distinction. 
To develop this analogy, I juxtapose a rule-based derivation (Heisenberg) with an Ehrenfest-style derivation (Schrödinger), tracing their epistemic asymmetry forward into the language of modern complexity theory and quantum computation.

Mara Beller’s \emph{Quantum Dialogue: The Making of a Revolution} portrays the birth of quantum mechanics as a fundamentally dialogical process—shaped by argument, persuasion, and interpretive negotiation among physicists \cite{Beller1999}. 
Her method is historiographical and rhetorical, reconstructing how meaning emerged through correspondence, debate, and controversy. 
In contrast, the present essay approaches the same episode structurally, treating the contrast between matrix and wave mechanics not as a social debate but as a formal opposition between two epistemic architectures.

Section \ref{Hist} examines the historical tensions that unfolded in early quantum mechanics between 1925 and 1932. 
Section \ref{Alg} explores the epistemic architectures implicit in their competing formalisms. 
Section \ref{Comp} translates this opposition into the language of computational complexity. 
Throughout, the contrast between matrix mechanics and wave mechanics is treated not merely as a historical dispute but as a structural opposition between modes of knowing.

\section{1925-1932: Matrix Mechanics vs. Wave Mechanics} \label{Hist}

\subsection{Matrix Mechanics}

\subsubsection{Bohr-Sommerfeld}

In Bohr’s picture of the atom, it can exist only in certain discrete, stationary states (or orbits) in which it does not radiate energy, even though classical electrodynamics predicts that it should. Each stationary state corresponds to a definite energy $E_n$:
\begin{equation} \label{EnEm}
E = E_n, \qquad n = 1, 2, 3, \ldots    
\end{equation}
Thus, when the electron is in one of the discrete stationary orbits (energy levels $E_n$) \eqref{EnEm}, the orbit is stable.  

However, when an atom makes a transition between two different stationary states $n$ and $m$ ($n\neq m$), it emits or absorbs radiation of frequency \cite{Bohr}:
\begin{equation}\label{BohrFreq-nu}
\nu_{nm}=\frac{E_n-E_m}{h},
\end{equation}
equivalently, in angular frequency:
\begin{equation}\label{BohrFreq-omega}
\omega_{nm}=\frac{E_n-E_m}{\hbar}.
\end{equation}

In the old quantum theory, for an electron in a periodic orbit, as in Bohr’s model, the position may be expressed as a Fourier series:
\begin{equation}\label{Fur}
x(t)=\sum_{\alpha=-\infty}^{\infty} a_{\alpha}(n)\,e^{i\,\omega_{\alpha}(n)\,t},
\end{equation}
with the velocity of the electron in its classical periodic orbit:
\begin{equation} \label{Fur-1}
\dot{x}(t)=\sum_{\alpha=-\infty}^{\infty} i\,\omega_{\alpha}(n)\,a_{\alpha}(n)\,e^{i\omega_{\alpha}(n)\,t},    
\end{equation}
and with corresponding momentum:
\begin{equation} \label{mom}
p(t) = m \dot{x}(t) 
= i m \sum_{\alpha=-\infty}^{\infty} \omega_{\alpha}(n)\,a_{\alpha}(n)\, e^{i \omega_{\alpha}(n) t}.
\end{equation}
The Fourier coefficients $a_{\alpha}(n)$ and frequencies $\omega_{\alpha}(n)$: 
\begin{equation} \label{orbit} 
a_{\alpha}(n), \qquad \omega_{\alpha}(n) = \alpha\, \omega(n), 
\end{equation}
depend on the orbit (quantum number) $n$, and satisfy $a_{-\alpha}(n)=a_{\alpha}(n)^{*}$ for real $x(t)$; that is, $x(t)$ represents a real physical coordinate, though complex exponentials are used for algebraic convenience.

Bohr introduced \eqref{Fur} to show that, in the limit of large quantum numbers, the frequencies of emitted radiation predicted by his two quantum postulates---namely, 
that atoms occupy discrete stationary states (or orbits) \eqref{EnEm} and emit radiation of frequency \eqref{BohrFreq-nu} during transitions--- coincide with the harmonics of the classical motion. This is an expression of the correspondence principle. He wrote \cite{Bohr1918}:
\begin{quote}
Now in a stationary state of a periodic system the displacement of the particles in any given direction may always be expressed by means of a Fourier–series as a sum of harmonic vibrations.
\end{quote}

Strictly speaking, equation \eqref{Fur} is semi-classical, bridging the gap between classical mechanics and quantum mechanics.

Arnold Sommerfeld introduced the quantization condition~\cite{Sommerfeld}:
\begin{equation}\label{oint}
\oint p\,dq = n\,h,
\end{equation}
according to which, each allowed orbit is characterized by a quantum number $n$ and a corresponding energy $E_n$.  
Thus, equation~\eqref{oint} selects the admissible periodic motions, while the Fourier series \eqref{Fur} describes the corresponding periodic trajectory 
for each allowed value of $n$.

\noindent Expressed discretely, the condition \eqref{oint} may be written, up to factors of $2\pi$, as: 
\begin{equation} \label{ointd}
\Delta\!\left( \oint p\,dq \right) = h.
\end{equation}
From the expressions \eqref{Fur}, \eqref{Fur-1}, \eqref{mom}, and \eqref{oint}, the action for this orbit is:
\begin{equation}
J(n)=\oint p\,dq =\int_{0}^{T(n)} p(t)\,\dot{x}(t)\,dt
= m\int_{0}^{T(n)} \dot{x}(t)^{2}\,dt.    
\end{equation}
The action variable can be obtained as:
\begin{equation} \label{momentum}
J(n) = 2\pi m \sum_{\alpha=-\infty}^{\infty} \alpha\, \omega_{\alpha}(n)\, |a_{\alpha}(n)|^2,
\end{equation}
where the sum runs over all integer harmonics $\alpha$.  

In the old quantum theory, this action variable was constrained by the quantization rule \eqref{oint}, thereby selecting discrete orbits from the continuous set of classical trajectories. Thus, only those periodic trajectories satisfying $J(n) = nh$ were allowed.

\subsubsection{Heisenberg}

Heisenberg’s 1925 paper “Quantum-theoretical Reinterpretation of Kinematic and Mechanical Relations” (\emph{Umdeutung} “reinterpretation”) marked the actual birth of quantum mechanics. Heisenberg retained the formal structure of classical mechanics but reinterpreted its symbols. The particle’s coordinates became arrays of complex numbers representing transitions between quantum states. These arrays obeyed a rule equivalent to matrix multiplication—though Heisenberg, unaware of matrix algebra, expressed it instead as frequency addition among transition amplitudes.
By replacing the unobservable Bohr–Sommerfeld orbits with algebraic relations among observable transitions, the \emph{Umdeutung} paper formulated a dynamics expressed entirely in terms of quantities labeled by the characteristic Bohr frequencies. As Heisenberg himself wrote, the older quantum rules “contain, as basic element, relationships between quantities that are apparently unobservable in principle, e.g., position and period of revolution of the electron. Thus, these rules lack an evident physical foundation” \cite{Heisenberg1925}.
His new mechanics, therefore, discarded the classical notion of the electron’s path and replaced it with relations among observable transition quantities only.

Heisenberg began from the Bohr--Sommerfeld semi-classical description of periodic motion, but made a radical departure. 
In the old quantum theory, the motion in a stationary orbit was represented by the Fourier series \eqref{Fur}, with coefficients $a_{\alpha}(n)$ \eqref{orbit} that depended on the orbit labeled by $n$. 
Heisenberg's key insight was to replace these \emph{unobservable orbit quantities} with \emph{observable transition quantities} between stationary states \cite{Heisenberg1925}:
\begin{equation} \label{transition}
X(n,n-\alpha),
\end{equation}
associated with the transition frequencies:
\begin{equation} \label{transition2}
\omega(n,n-\alpha) = \frac{E_n - E_{n-\alpha}}{\hbar}.
\end{equation}
corresponding to the emitted or absorbed radiation:
\begin{equation}
X(n,n-\alpha)\, e^{i\,\omega(n,n-\alpha)\,t}.
\end{equation}
The substitution of orbit-based quantities by transition quantities transformed the semi-classical scheme into proper quantum mechanics.

Heisenberg’s goal was to reformulate Sommerfeld’s quantization condition \eqref{oint} in terms of 
observable quantities \eqref{transition} and \eqref{transition2}, rather than unobservable orbits $x(t)$ and $p(t)$ . 
Since the old quantum theory required $J = n h$, he rewrote the action \eqref{momentum} in discrete form as \cite{Heisenberg1925}:
\begin{equation} \label{quantization}
h = 4 \pi m \sum_{\alpha} \alpha\, \omega_{\alpha}\, |a_{\alpha}|^2,
\end{equation}
a relation connecting the amplitudes and frequencies to Planck’s constant $h$. The factor of $4$ (instead of $2$) arises from taking the difference between neighboring quantized actions $J_n - J_{n-1}$.
Equation~\eqref{quantization} thus expresses the quantization of action in terms of measurable Fourier amplitudes and frequencies.

In his \emph{Umdeutung} paper, Heisenberg eliminated all reference to unobservable electron orbits. 
While the trajectory $x(t)$ cannot be observed, the radiation emitted in transitions between stationary states \emph{can}. 
Accordingly, Heisenberg replaced the classical function $x(t)$ by a complete system of 
quantities describing transition amplitudes between stationary states $n$ and $m$:
\begin{equation} \label{TrAm}
x_{nm} \equiv X(n,m).
\end{equation}
The quantity describing the transition from state $m$ to $n$ can be written either as 
$X(n,m)$ or $x_{nm}$. In particular, for transitions between neighboring states differing by 
an integer step $\alpha$ in the quantum number ($m = n - \alpha$), Heisenberg wrote \cite{Heisenberg1925}:
\begin{equation} 
x_{nm} = X(n,n-\alpha) = X_{n,n-\alpha}.
\end{equation}
The transition~\eqref{transition} is therefore a special case of \eqref{TrAm}. 

\noindent The corresponding momentum amplitudes follow from the relation $p(t) = m\,\dot{x}(t)$:
\begin{equation} \label{pnm}
p_{nm} = i\,m\,\omega(n,m)\,x_{nm}.
\end{equation}
Hence, $x_{nm}$ and $p_{nm}$ represent the amplitudes of emission or absorption at frequency:
\begin{equation}
\omega_{nm} = \frac{E_n - E_m}{\hbar}.
\end{equation}
The full motion of the electron was thereby represented not by a single function of time but by an entire \emph{array of quantities}:
\begin{equation} \label{trans}
X(n,m)\, e^{i \omega(n,m) t}, \qquad \text{for all } n,m,
\end{equation}
later recognized by Born and Jordan as the elements of \emph{a matrix}. 

In Heisenberg’s formulation, the quantities $X_{nm}$ \eqref{TrAm} describe observable transition amplitudes whose squared magnitudes are proportional to the intensities of spectral lines:
\begin{equation} \label{Inm}
|X_{nm}|^2 \propto I_{nm},
\end{equation}
where $I_{nm}$ is the intensity of the emitted or absorbed radiation at the frequency $\omega_{nm}$. 

\noindent Only later, under Born’s 1926 statistical interpretation of Schrödinger’s wavefunction \cite{Born1, Born2}, were such quantities reinterpreted as probability amplitudes rather than radiation amplitudes.

Heisenberg found that, instead of continuous functions, one must use quantities depending on pairs of states, which combine according to the rule:
\begin{equation} \label{combine}
(xy)_{nm} = \sum_k x_{nk} y_{km}.
\end{equation}
This is precisely the law of matrix multiplication, although Heisenberg did not yet recognize it as such. 

In classical mechanics, quantities such as $x$ and $p$ are real-valued dynamical variables, continuous functions of time or of phase-space coordinates. Their multiplication is ordinary arithmetic. Thus, they commute under multiplication:
\begin{equation}
x p = p x,
\end{equation}
But in Heisenberg’s framework, each observable corresponds to a table of transition amplitudes between states. So, how do we form quantities like $xp$ and $px$ when $x$ and $p$ are not single values, but arrays of transitions? 
Using the transition amplitudes \eqref{TrAm} and 
\eqref{pnm} and the combination rule \eqref{combine}, Heisenberg defined the product 
of two quantities (such as $x$ and $p$) by summing over intermediate states \cite{Heisenberg1925}:
\begin{equation} \label{commutator}
(xp)_{nm} = \sum_k x_{nk} p_{km}, \qquad 
(px)_{nm} = \sum_k p_{nk} x_{km}.
\end{equation}
This composition rule combines the amplitude relations for $x_{nm}$ and $p_{nm}$ 
and marks the first appearance of the noncommutative multiplication law later recognized as matrix multiplication.
This insight, clarified by Born and Jordan, became the algebraic foundation of matrix mechanics.

\subsubsection{Born-Jordan}

Max Born and Pascual Jordan soon recognized Heisenberg's conceptual leap as the foundation of matrix mechanics \cite{Heisenberg1925, DuncanJanssen}.
In 1925, Born and Jordan recast Heisenberg’s arrays as genuine \emph{matrices}, introduced the formalism of matrix multiplication, wrote down the canonical commutation relation \eqref{sharpen} in quantum form, and—together with Heisenberg—systematized and extended this matrix mechanics \cite{Born-Jordan, Born-Heisen-Jordan}. Born immediately recognized that Heisenberg’s tables obeyed the rules of matrix multiplication and, with Jordan, reformulated them as a consistent algebraic system.

In their 1925 paper, Born and Jordan introduce the quantity \cite{Born-Jordan}:
\begin{equation} \label{pqqp}
d = pq - qp.
\end{equation}
They first noticed that $pq \neq qp$, i.e., that the order of multiplication matters when $p$ and $q$ are represented as matrices (\emph{non-commutativity}).

Born and Jordan reinterpreted the quantization rule \eqref{ointd} within the framework of Heisenberg’s matrix mechanics.
They represented the canonical variables $p$ and $q$ as infinite matrices $(p_{mn})$ and $(q_{mn})$, 
whose elements describe transitions between stationary states $m \to n$.  
The classical integral $\oint p\,dq$ over one period corresponds, in this scheme, to a sum over transitions that return the system to the same state $n$, i.e., to the diagonal components of the commutator matrix $(pq - qp)$.

Born and Jordan found that expressing the quantum condition \eqref{oint} in matrix form constrains precisely these diagonal components \cite{Born-Jordan}:
\begin{equation} \label{sharpen1}
(pq - qp)_{nn} = \frac{h}{2\pi i}.
\end{equation}
They then postulated that this relation should hold not only for the diagonal components but for \emph{all} matrix elements. They called~\eqref{sharpen1} the 
\emph{verschärfte Quantenbedingung}, the “sharpened quantum condition.”  
Thus, the Bohr–Sommerfeld quantization rule becomes, in matrix form \cite{Born-Jordan}:
\begin{equation} \label{sharpen}
pq - qp = \frac{h}{2\pi i}\,1,
\end{equation}
where $1$ denotes the identity matrix.  
This is the canonical commutation relation: the commutator of position and momentum is proportional to the identity operator.

Born and Jordan required that Heisenberg’s system of transition amplitudes \eqref{TrAm} $X(n,m)$ reproduce the classical relation between $p$ and $q$ for large quantum numbers 
(the correspondence principle). Starting from the Fourier coefficients that relate to $x_{nm}$ and $p_{nm}$, they used Heisenberg’s \emph{quantization condition} \eqref{quantization}, expressed $p$ and $x$ in terms of these Fourier components:
\begin{equation}
p_{nm} \propto i m \omega_{nm} x_{nm},
\end{equation}
and obtained:
\begin{equation}
\sum_m (p_{nm} x_{mn} - x_{nm} p_{mn}) = \frac{h}{2\pi i}.
\end{equation}
Recognizing that the left-hand side is the matrix commutator $[p,x]$, they generalized this to the operator equation~\eqref{sharpen}.

Heisenberg’s amplitudes $x_{nm}$ \eqref{TrAm} thus behave algebraically like matrix elements.  
Born and Jordan represented both position and momentum as systems of transition amplitudes:
\begin{equation} 
x(t) = \{ x_{nm} e^{i \omega_{nm} t} \}, \qquad 
p(t) = \{ p_{nm} e^{i \omega_{nm} t} \}.
\end{equation}
Using Heisenberg’s combination rule \eqref{combine} or \eqref{commutator} \cite{Heisenberg1925}:
\begin{equation}
(xp)_{nm} = \sum_k x_{nk} p_{km},
\end{equation}
they showed that, in general:
\begin{equation}
xp \neq px,
\end{equation}
and that the commutator between these matrices yields the quantization condition proportional to $\dfrac{h}{2\pi i}$.

In Heisenberg’s original formulation, the quantities $x_{nm}$ represented observable transition amplitudes whose squared magnitudes are proportional to the intensities 
of spectral lines. Using the commutation relation \eqref{sharpen}, Born and Jordan differentiated and generalized the result to arbitrary functions $H(p,q)$, obtaining for any dynamical quantity $g(p,q)$~\cite{Born-Jordan}:
\begin{equation} \label{GH}
\dot{g} = \frac{2\pi i}{h}\,(H g - g H).
\end{equation}
Equation \eqref{GH} is the \emph{matrix-mechanical equation of motion}, the Heisenberg form of Hamilton’s equations.  
In classical mechanics, Hamilton’s equations specify the time evolution of any quantity $g$:
\begin{equation}
\dot{g} = \{ g, H \} = \frac{\partial g}{\partial q}\frac{\partial H}{\partial p} - \frac{\partial g}{\partial p}\frac{\partial H}{\partial q}.
\end{equation}
Born and Jordan extended this to the quantum regime by replacing the Poisson bracket with the commutator, scaled by $\dfrac{2\pi i}{h}$:
\begin{equation}
\{ g, H \} \;\longrightarrow\; \frac{2\pi i}{h}\,[H,g].
\end{equation}
Hence, Eq. \eqref{GH} expresses that the time evolution of any observable $g$ is determined by its commutator with the Hamiltonian $H$.  
The commutator thus replaces the classical Poisson bracket, and the canonical relation \eqref{sharpen} 
embodies the structure that the old quantum condition \eqref{oint} had previously represented.  
Born and Jordan’s insight completed the transition from the Bohr–Sommerfeld theory to the operator formalism of quantum mechanics.

\subsubsection{Born-Heisenberg-Jordan}

In the following year, Born, Heisenberg, and Jordan extended these ideas into a systematic operator calculus. In 1926, their \emph{Dreimännerarbeit} paper introduced an operator calculus that already foreshadowed the structure of a Hilbert space, albeit in an intuitive manner.

Born, Heisenberg, and Jordan therefore identified \emph{Hermitian matrices as the mathematical representatives of physical observables}. This choice ensured that the algebraic structure of quantum mechanics automatically produced measurement results consistent with physical experience. It was a conceptual breakthrough because the formal properties of matrices now mirrored the empirical properties of measurement. They showed that a Hermitian matrix $a = (a_{mn})$ equals its Hermitian conjugate (complex conjugate transpose), $a = a^\ast = (a_{nm}^\ast)$, meaning that each element in the $m$th row and $n$th column is the complex conjugate of the element in the $n$th row and $m$th column. The associated \emph{quadratic form}—quadratic because it involves products like $x_m x_n^{\ast}$—is \cite{Born-Heisen-Jordan}:
\begin{equation} \label{quad}
A(x, x^{\ast}) = \sum_{m,n} a_{mn}\, x_m x_n^{\ast}.
\end{equation}
If $a_{mn}$ is Hermitian, then $A(x,x^{\ast})$ is always real. In physical terms, this quadratic form corresponds to the \emph{expectation value} of an observable. Thus, Hermiticity guarantees that all expectation values, and hence all measurable quantities, are real \cite{Born-Heisen-Jordan}.

A \emph{Hilbert space} is a complete complex vector space equipped with an inner product that defines lengths, angles, and projections—and, in quantum mechanics, expectation values. When Born, Heisenberg, and Jordan used expressions like \eqref{quad}, they were already operating with the same algebraic framework as a Hilbert space, just without the later terminology. Vectors $x$ represent states, Hermitian matrices represent observables, and inner products such as \eqref{quad} represent measurable averages. Because the mathematical machinery for infinite-dimensional spaces had not yet been fully formalized in the physics literature, they worked with finite matrices instead.

Born, Heisenberg, and Jordan explicitly cited the mathematical work of David Hilbert and Ernst Hellinger (1904–1910) on infinite-dimensional quadratic forms as the theoretical foundation for their Hermitian formalism—an early precursor to what would become the theory of Hilbert spaces. They recognized that their quantum observables (matrices representing momentum, energy, and so forth) were \emph{unbounded}, meaning they could have arbitrarily large eigenvalues. Nevertheless, they assumed that the Hilbert–Hellinger results for bounded Hermitian forms would also hold for these unbounded cases \cite{Born-Heisen-Jordan}. That assumption was a bold extrapolation because they extended a mathematical theory on physical grounds, trusting the internal consistency of the new mechanics.

Soon afterward, in 1927, John von Neumann would rigorously prove that every self-adjoint operator, including unbounded ones, on a Hilbert space admits a spectral decomposition \cite{Neumann1927}. Born, Heisenberg, and Jordan had already formulated the finite-dimensional version of this theorem in 1926, expressing any Hermitian matrix as a unitary transformation to diagonal form and even anticipating the possibility of a continuous spectrum \cite{Born-Heisen-Jordan}. Von Neumann’s proof extended their physical and algebraic insight to the general case of infinite-dimensional operators, thereby giving complete mathematical legitimacy to the Born–Heisenberg–Jordan framework and completing the formal foundations of quantum mechanics.

In essence, their paper speaks the language of matrices but already \emph{operates conceptually in the realm of linear operators on complex vector spaces}. It invokes spectral decomposition, orthogonality, Hermiticity, and unitarity—the fundamental pillars of Hilbert-space theory—and explicitly recognizes the need for a theory of unbounded operators and continuous spectra \cite{Born-Heisen-Jordan}. Von Neumann’s 1927–1932 \emph{Mathematische Grundlagen der Quantenmechanik} would supply the missing rigor, turning the intuitive Hilbert–Hellinger framework they had assumed into a complete mathematical edifice \cite{Neumann1927, Neumann1932}.

In short, Heisenberg provided the physical reinterpretation, Born and Jordan supplied the algebraic formalism, and von Neumann later secured the mathematical foundation.

\subsubsection{Dirac}

Almost simultaneously, and independently in formulation, Paul Dirac reformulated Heisenberg’s matrix mechanics into a general algebra of quantum quantities. He introduced rules for the addition, multiplication, and differentiation of noncommuting quantum quantities $x, y$ with components $(x_{mn})$. He identified the correspondence between the quantum commutator and the classical Poisson bracket \cite{Dirac-1, Dirac-2}:
\begin{equation}
xy - yx = \frac{i h}{2\pi}\,\{x,y\},
\end{equation}
where $\{x,y\}$ denotes the classical Poisson bracket. He also expressed time evolution as:
\begin{equation}
\dot{x} = \frac{i}{\hbar}[H,x] \equiv \frac{i}{\hbar}(H x - x H),
\end{equation}
thus anticipating the Heisenberg equation of motion in operator language~\eqref{Heis1} \cite{Dirac-1,Dirac-2}. Taken together, these developments constitute what is now called the \emph{Heisenberg picture}.

Helge Kragh highlights that while Dirac’s paper, "The Fundamental Equations of Quantum Mechanics" \cite{Dirac-1} was completed independently in Britain, it coincided almost precisely with the Born-Jordan formulation developed in Germany. Both introduced matrix representation and the commutation relation \eqref{sharpen}. But Kragh emphasizes that Dirac’s version uniquely linked it to the classical Poisson bracket, giving it deeper physical meaning and formal elegance \cite{Kragh}.

According to Born and Jordan, the noncommutativity $\mathbf{pq} - \mathbf{qp} \neq 0$ embodies the canonical structure that in classical theory is expressed by the Poisson bracket. Since $\dfrac{1}{i}=-i$, we can rewrite \eqref{sharpen} as
\begin{equation} \label{qp1}
pq-qp = -\,i\,\frac{h}{2\pi}.
\end{equation}
Equivalently:
\begin{equation} \label{qp}
\{q,p\} = 1 \;\longrightarrow\; [\mathbf{q},\mathbf{p}] = i\,\frac{h}{2\pi}.
\end{equation}
Equation \eqref{qp} uses Dirac's notation, i.e., the commutator is defined as $[\mathbf{q},\mathbf{p}]=\mathbf{q}\mathbf{p}-\mathbf{p}\mathbf{q}$. With $\hbar = \dfrac{h}{2\pi}$, this becomes:
\begin{equation} \label{can}
[\mathbf{q},\mathbf{p}] = i\hbar \qquad \text{or:} \qquad [Q, P] = i\hbar.
\end{equation}
It represents the canonical commutation relation in its standard quantum form.

Born and Jordan derived from \eqref{sharpen} (equivalently, from \eqref{qp}) the matrix equations of motion \eqref{GH}. In Dirac's notation, their equation reads \cite{Born-Jordan}:
\begin{equation}
\frac{d g}{d t} = \frac{2\pi i}{h}\,[H, g].
\end{equation}
This is the quantum analogue of Hamilton’s equations and anticipates the Heisenberg equation of motion in operator language:
\begin{equation} \label{Heis1}
\frac{dA}{dt} = \frac{i}{\hbar}\,[H, A].  
\end{equation}

\subsubsection{Heisenberg} \label{HER}

In 1927, Heisenberg published his seminal paper “Über den anschaulichen Inhalt der quantentheoretischen Kinematik und Mechanik” on the uncertainty principle \cite{Heisenberg1927}. In this work, he cited Max Born for their joint development of matrix mechanics with Pascual Jordan and credited Jordan specifically for the Dirac–Jordan transformation theory. He acknowledged Wolfgang Pauli for his influence, discussions, and early formulation of related ideas, thanking him explicitly in a footnote. Paul Dirac is mentioned repeatedly, particularly in connection with transformation theory and the commutation relations, as Heisenberg directly linked his own results to Dirac’s formalism. Heisenberg treats Paul Ehrenfest’s work as foundational and insightful, building upon it. Louis de Broglie appears in a positive light, associated with wave–particle duality and the role of de Broglie wavelengths in defining the limits of spatial and temporal measurements. Albert Einstein is cited respectfully for his contributions to the photoelectric effect and his insights into the wave-particle duality. Niels Bohr, mentioned in the footnotes as a source of inspiration, is credited for formative conversations on the limits of classical visualization; Heisenberg explicitly notes that the paper evolved from ideas developed in collaboration with Bohr \cite{Heisenberg1927}.

In striking contrast, Heisenberg mentioned Erwin Schrödinger, citing his 1926 article on the transition from micro- to macromechanics—only to dismiss it as inadequate. He regarded Schrödinger’s wave-mechanical reasoning as fundamentally misguided, insisting that the correspondence with classical mechanics follows not from wave packets but from the operator formalism itself. Heisenberg acknowledged that Schrödinger had called quantum mechanics a formal theory of repellent abstraction and lack of visualizability. While Heisenberg agrees that Schrödinger’s wave mechanics achieved a mathematical and partly intuitive (and thus “in a sense” visualizable) formulation of the quantum laws, he insists that its popular visualizability led physics down the wrong path. In Heisenberg’s view, the appeal of wave mechanics to pictures and classical imagery distracts from the deeper, non-visual, and abstract nature of quantum phenomena. He argues that the genuine progress of quantum mechanics—emerging from the works of Bohr and the new matrix formulation—lies not in making it visually intuitive, but in rigorously expressing relations between observable quantities. Thus, Heisenberg rejects Schrödinger’s philosophical orientation toward classical “Anschaulichkeit” (visualizability) \cite{Heisenberg1927}.
By 1927, Heisenberg’s resistance to wave mechanics was complete. In 1926–1927, a fierce intellectual battle unfolded between Schrödinger’s wave mechanics and Heisenberg’s matrix mechanics \cite{Beller1999}.

As discussed earlier, Born, Jordan, and Heisenberg developed Heisenberg’s theory into a systematic formalism \cite{Born-Jordan, Born-Heisen-Jordan}. In the Born-Jordan paper, Born and Jordan wrote the "sharpened quantum condition" \eqref{sharpen}. This was the operator-matrix version of the old Bohr–Sommerfeld quantization rule \eqref{oint}. It expressed the same physical constraint, but now in a fully non-commutative matrix algebraic form. In matrix mechanics, $q$ and $p$ were not numbers but infinite matrices representing observables. Their commutator is proportional to $\dfrac{h}{2 \pi i}$. 

In his 1927 paper on the uncertainty principle, Heisenberg wrote essentially the same relation \eqref{sharpen} \cite{Heisenberg1927}:

\begin{equation} \label{sharpen2}
pq - qp = \frac{h}{2\pi i},
\end{equation}
but his intent was different.
He was not deriving it as a mathematical quantization rule, but rather reinterpreting it physically as the foundation of indeterminacy between conjugate quantities.

So while Born and Jordan introduced this commutation rule as a formal algebraic postulate—a condition ensuring the correct quantum energy levels—Heisenberg now gave it a physical meaning. The relation \eqref{sharpen2} expresses the impossibility of simultaneously defining both $p$ and $q$ precisely.
This shift — from a rule of quantization to a statement of physical limitation — is what transforms the “sharpened quantum condition” \eqref{sharpen} into the \emph{Uncertainty Principle}. 
What seemed like a small mathematical reinterpretation was, in fact, a philosophical earthquake. Heisenberg took what had been a formal algebraic postulate (Born–Jordan’s sharpened condition \eqref{sharpen}) and redefined it as a limit on what can be known about nature. This changed the very epistemology of physics, from describing what \emph{is} to explaining what can be observed and predicted.

Heisenberg introduced the $\gamma$-microscope as a thought experiment intended to illustrate the reciprocal limitations involved in measuring the position and momentum of an electron. The idea is that to localize an electron
very precisely, one must illuminate it with radiation of extremely short wavelength, such as $\gamma$-rays. Short wavelength improves the spatial resolution $\Delta x$, but $\gamma$-photons carry large momentum; when a photon
is scattered into the microscope, it imparts an unpredictable momentum kick to the electron. Thus, decreasing $\Delta x$ by using shorter wavelengths
necessarily increases the disturbance in momentum $\Delta p$. Heisenberg used this to motivate the uncertainty relation. In his 13 July 1935 letter to Einstein, Schrödinger criticizes this device, arguing
that a $\gamma$-microscope does not in fact measure any variable that possesses a true canonical conjugate partner in the operator sense of equation \eqref{sharpen2} \cite{Meyenn} (see the discussion in section \ref{Schrodinger1}).

When Heisenberg, working with Born and Jordan in Göttingen, formulated matrix mechanics in 1925, the theory was so radically abstract that even its creators struggled to interpret it. The very word “matrix” was unfamiliar to most physicists \cite{Condon}. As Max Jammer later observed, many either “did not know what matrices were or, if they did, were reluctant to apply them to theoretical problems” \cite{Jammer}. 
Born, eager to put the new mechanics on a solid mathematical footing, searched for an assistant familiar with both physics and the abstract machinery of linear algebra. He first approached Pauli, who declined the offer, but soon made a significant contribution to the new theory. Few others seemed qualified—or willing—to work with such strange mathematics.

In this intellectual climate, Born and Heisenberg sought advice from David Hilbert, the legendary mathematician whose work had already revolutionized the foundations of mathematics and whose influence permeated Göttingen. The physicists explained that their new formulation relied on arrays of numbers, matrices, whose manipulation was proving devilishly difficult.

Hilbert, whose intuition about mathematics often preceded its formalization, remarked that matrices had appeared naturally in his own work when solving boundary-value problems in differential equations. In those cases, the matrix elements represented relationships between wave-like functions, and their eigenvalues corresponded to measurable quantities, such as frequencies or energies. He suggested, with characteristic simplicity, that they might try to find the differential equation whose eigenvalues generated those matrices—in essence, to look for the wave picture behind the algebraic symbols.

To Born and Heisenberg, however, this sounded like a retreat into classical imagery. Their matrix mechanics was meant to abandon continuous pictures entirely; to invoke a differential equation seemed to miss the point. They politely dismissed Hilbert’s idea.

Only a few months later, Erwin Schrödinger derived a wave equation that described the same quantized energy levels as Heisenberg’s matrices. 
When Hilbert heard of Schrödinger’s discovery, he could not resist a smile. As Edward Condon recalled, the mathematician had "a great laugh" on Born and Heisenberg, remarking that they could have anticipated Schrödinger’s breakthrough six months earlier if only they had listened to him \cite{Condon}.

The irony was profound. Hilbert’s intuition—rooted purely in mathematical reasoning rather than physical imagery—had led him straight to the missing link between matrix mechanics and wave mechanics. The physicists, convinced they were venturing into wholly new territory, had dismissed the very idea that bridged their abstract matrices and Schrödinger’s continuous waves.
Hilbert’s laughter was not unkind; it was the laughter of someone who had glimpsed the underlying harmony long before others could see it.

Schrödinger’s wave mechanics emerged in 1926 as a unified formalism for quantum mechanics. Still, it, too, was the culmination of a collective evolution of ideas. It was further developed and refined by others after his series of four seminal papers “Quantization as an eigenvalue problem” \cite{Schrod1, Schrod2, Schrod3, Schrod4}, and the fifth paper \cite{Schrod5}. 

\subsection{Wave Mechanics}

\subsubsection{De Broglie and Einstein}

Several key strands were already in place before Schrödinger entered the scene. In the years 1923–1924, Louis de Broglie proposed the wave–particle duality of matter, where classical mechanics is expressed as a wave-like differential equation for the action $S$, which influenced Schrödinger to develop quantum waves. Albert Einstein, in his seminal works, connected quantized energy transitions and radiation statistics, establishing the conceptual need for a wave description of quanta. 

In his PhD dissertation published in 1925, de Broglie starts from Max Planck's relation:
\begin{equation} \label{Planck}
E_0 = h \nu_0,    
\end{equation}
and applies it to rest energy, writing the relation \cite{Broglie}:
\begin{equation} \label{35}
m_0c^2 = h\nu_0,
\end{equation}
assigning a frequency $\nu_0$ to the rest energy $E_0 = m_0c^2$.
This is de Broglie’s energy–frequency equivalence, formally analogous to Planck’s relation for light.

\noindent For a moving particle (in the laboratory frame), the total energy is:
\begin{equation} \label{EMC}
E = \gamma m_0 c^2 = \frac{m_0c^2}{\sqrt{1 - v^2/c^2}},
\end{equation}
and assuming the same proportionality $E = h\nu$, de Broglie obtains \cite{Broglie}:
\begin{equation} \label{36}
\nu = \frac{\gamma m_0 c^2}{h}, 
\qquad 
\gamma = \frac{1}{\sqrt{1 - v^2/c^2}}.
\end{equation}
For a free particle (laboratory frame), \eqref{EMC} and:
\begin{equation} \label{pmv}
p = \gamma m_0v,    
\end{equation}
hold.

\noindent De Broglie defines the phase velocity $V$ of the accompanying "phase wave":
\begin{equation} \label{lamnu}
V = \lambda \nu.    
\end{equation}
He combines the relations \eqref{lamnu} and:
\begin{equation} \label{broglie1}
\nu = \frac{E}{h} \quad \text{\eqref{36}}.    
\end{equation}
Since for a free particle:
\begin{equation} \label{VEP}
V = \dfrac{E}{p},    
\end{equation}
with $E$ given by \eqref{EMC} and $p$ by \eqref{pmv}, we get \cite{Broglie}:
\begin{equation} \label{Vnu}
\lambda=\frac{V}{\nu} = \dfrac{\frac{E}{p}}{\frac{E}{h}} = \dfrac{h}{p}.  
\end{equation}
Thus, a particle of momentum $p$ has a wavelength:

\begin{equation} \label{broglie}
p= \dfrac{h}{\lambda}.    
\end{equation}
The relations \eqref{broglie1} and \eqref{broglie} connect the particle’s energy and momentum with the frequency and wavelength of an associated wave.

In sections \S 8-\S 9 of his 1925 paper on Bose--Einstein statistics, Einstein proposed that every material particle is accompanied by an undulatory field (a "Gespensterfeld", “ghost field”), analogous to the electromagnetic field accompanying a photon. This was the first clear statement of matter waves, a year before Schrödinger formulated wave mechanics \cite{Einstein1925}. 
In section \S8 of his 1925 paper, Einstein, responding to de Broglie’s 1924--1925 hypothesis of matter waves, proposed that every material particle is accompanied by an undulatory “ghost field” (\emph{Gespensterfeld}). Einstein derived an expression that de Broglie had already obtained \cite{Broglie}. He divided the relativistic expressions for energy \eqref{EMC} and momentum \eqref{pmv}, and interpreted their ratio physically as the \emph{phase velocity} $V$ of the accompanying field \cite{Einstein1925} [equation \eqref{VEP}]:
\begin{equation} \label{37}
V = \frac{E}{p} = \frac{c^2}{v},
\end{equation}
which, being superluminal, represents a non-energetic phase propagation rather than a transport of energy.

Einstein then considered the non-relativistic regime ($v \ll c$), characterized by small velocities compared to the speed of light, $c$. Neglecting the Lorentz factor $\gamma$ ($\gamma \approx 1$) in~\eqref{pmv}, and substituting $p = mv$ into~\eqref{Vnu}, he obtained for the accompanying undulatory field:
\begin{equation} \label{38}
\lambda = \frac{V}{\nu} = \frac{h}{mv}.
\end{equation}
This relation expresses the wave--particle duality for matter. Every moving material particle is accompanied by an undulatory “ghost field,” whose phase velocity $V = \dfrac{c^2}{v}$ exceeds the speed of light. At the same time, the transport of energy occurs with the group velocity, equal to the particle’s actual velocity $v$.

\subsubsection{Schrödinger}

In section \S 2 of his second paper, "Geometrical and Undulatory Mechanics", Schrödinger transformed the Einstein--de Broglie picture into a continuous mechanics. Einstein’s "ghost field," which he had introduced in 1925 as an undulatory accompaniment to every material particle, became in Schrödinger’s hands a mathematically real wave in configuration space ($q$-space). 

Schrödinger retained de Broglie-Einstein’s energy--frequency and wavelength--momentum correspondences \eqref{broglie1} and \eqref{broglie}, which had already suggested that matter possesses wave-like properties. At the same time, he incorporated de Broglie’s distinction between phase and group velocities, recognizing that while the phase wave can propagate faster than light, the group velocity---the speed of a wave packet---corresponds to the actual motion of the particle \cite{Schrod2}.

We begin from the de Broglie–Einstein correspondences \eqref{broglie1} and \eqref{broglie}.
In classical mechanics, the motion of a particle in a potential $V(q)$ is governed by the Hamilton--Jacobi equation:
\begin{equation} \label{HJ}
\frac{\partial S}{\partial t} + \frac{1}{2m}(\nabla S)^2 + V(q) = 0,
\end{equation}
where $S(q,t)$ is Hamilton’s principal function (the classical action). 

By embedding the correspondences \eqref{broglie1} and \eqref{broglie} into the Hamilton--Jacobi formalism of classical mechanics, Schrödinger reinterpreted the classical action $S(q,t)$ as the phase of a wave function.
Schrödinger’s idea was to reinterpret $S(q,t)$ as the \emph{phase} of an accompanying wave, setting:
\begin{equation} \label{psi}
\psi = e^{\frac{iS}{\hbar}}.
\end{equation}
Once Schrödinger adopts \eqref{psi} within his WKB (Wentzel–Kramers–Brillouin)-type derivation, he is already using de Broglie-Einstein’s energy--frequency and wavelength--momentum correspondences \eqref{broglie1} and \eqref{broglie} in differential form:
\begin{equation} \label{EinBrog}
\nabla S = p, \qquad -\,\frac{\partial S}{\partial t} = E,
\qquad
k = \frac{p}{\hbar}, \qquad \omega = \frac{E}{\hbar}.
\end{equation}
Differentiating \eqref{psi} gives:

\noindent 1. Time derivative:
\begin{equation} \label{d1}
\frac{\partial \psi}{\partial t}
= \frac{\partial}{\partial t}\!\left(e^{\frac{iS}{\hbar}}\right)
= e^{\frac{iS}{\hbar}} \cdot \frac{i}{\hbar}\frac{\partial S}{\partial t}
= \frac{i}{\hbar}\frac{\partial S}{\partial t}\,\psi.
\end{equation}
Substituting the Einstein relation, we get:
\begin{equation} \label{d11}
\frac{\partial \psi}{\partial t}
= - i \,\omega \, \psi.
\end{equation}
2. Gradient:
\begin{equation} \label{grad}
\nabla \psi
= \nabla\!\left(e^{\frac{iS}{\hbar}}\right)
= e^{\frac{iS}{\hbar}} \cdot \frac{i}{\hbar}\nabla S
= \frac{i}{\hbar}(\nabla S)\,\psi.    
\end{equation}
Substituting the de Broglie relation \eqref{EinBrog}, yields:
\begin{equation} \label{grad1}
\nabla \psi
= \frac{i}{\hbar}(\nabla S) \psi \, = \, i k \psi.    
\end{equation}
3. Laplacian. Starting from the gradient result, we take the divergence and use the product rule:
\begin{equation} \label{d2}
\nabla^2 \psi
= \nabla \cdot \left(\frac{i}{\hbar}\psi\,\nabla S\right)
= \frac{i}{\hbar}\Big[(\nabla \psi)\!\cdot\!(\nabla S) + \psi\,\nabla^2 S\Big].    
\end{equation}
Substituting \eqref{grad}, gives:
\begin{equation}
\nabla^2 \psi
= \frac{i}{\hbar}\left[\frac{i}{\hbar}(\nabla S)^2\psi + \psi\,\nabla^2 S\right]
= \left[-\frac{1}{\hbar^2}(\nabla S)^2 + \frac{i}{\hbar}\nabla^2 S\right]\psi.    
\end{equation}
To turn a real equation (for $S$) into a complex one (for $\psi$), we multiply the Hamilton--Jacobi equation \eqref{HJ} by $\dfrac{i}{\hbar}$: 
\begin{equation} \label{HJ1}
\dfrac{i}{\hbar}\frac{\partial S}{\partial t} + \dfrac{i}{\hbar}\frac{(\nabla S)^2}{2m} + \dfrac{i}{\hbar}V = 0.
\end{equation}
Now the derivatives of $S$ can be replaced by derivatives of $\psi$ using the relations \eqref{d1}, \eqref{grad}, and \eqref{d2}. Multiplying \eqref{HJ} by $\dfrac{i}{\hbar}$ shows that using \eqref{d1}, the first term becomes proportional to $\dfrac{\partial \psi}{\partial t}$. Thus, we replace each term in the multiplied Hamilton–Jacobi equation \eqref{HJ1} using the derivative relations \eqref{d1}, \eqref{grad}, and \eqref{d2}. 
This gives, after rearranging and multiplying by $i \hbar$ an intermediate “almost-Schrödinger” equation with the extra term:
\begin{align} \label{waveeq}
i\hbar\,\frac{\partial \psi}{\partial t}
&= \left[-\frac{\hbar^2}{2m}\nabla^2 + V(q)
+ \frac{i\hbar}{2m}\nabla^2 S\right]\psi.
\end{align}
Thus, by embedding the de Broglie–Einstein relations \eqref{EinBrog} into the Hamilton--Jacobi formalism and identifying the action $S$ with the phase of a complex wave amplitude \eqref{psi}, Schrödinger converted the classical trajectory equation into a wave equation \eqref{waveeq}.

\noindent If we now relax the assumption that $S$ is a real classical function and allow it to acquire an imaginary part that accounts for a slowly varying amplitude $A(q,t)$, writing $\psi = A\,e^{\frac{iS}{\hbar}}$, the additional term $\dfrac{i\hbar}{2m}\nabla^2 S$ combines with the amplitude derivatives to produce the full linear Schrödinger equation:

\begin{equation} \label{Schrodinger}
i\hbar\,\frac{\partial \psi}{\partial t} 
= 
\left[-\frac{\hbar^2}{2m}\nabla^2 + V(q)\right]\psi,
\end{equation}
the time-dependent Schrödinger equation.
The result is a continuous, differential expression of the discrete heuristic ideas or relations \eqref{EinBrog} proposed by Einstein and de Broglie \cite{Schrod2}.

In this way, Schrödinger replaced Einstein’s qualitative "ghost field" with a quantitative, dynamical field that could predict atomic spectra and interference phenomena. The wave function thus became the precise mathematical realization of the undulatory principle that Einstein had intuited and de Broglie had formulated, transforming the speculative wave--particle duality into a complete and predictive physical theory.

Matrix mechanics was introduced in 1925–1926 as an abstract, algebraic formulation of quantum theory. Schrödinger reinterpreted the new quantum mechanics in terms of a continuous wave equation, providing an intuitively visual counterpart to the abstract approach of matrix mechanics.  
In his \emph{Annalen der Physik} papers of that year \cite{Schrod1, Schrod2, Schrod3, Schrod4}, Schrödinger’s approach remained analytic and differential, not operator-theoretic. He introduced wave mechanics through partial differential equations rather than through the later operator formalism.

In Section 3 of his second paper, Schrödinger applies the newly formulated wave equation to specific systems. The first system he chose to test against Heisenberg’s matrix mechanics is the Planck oscillator, i.e., the quantum harmonic oscillator \cite{Schrod2}. 
Schrödinger begins by writing his stationary wave equation that describes how standing waves form in a bound potential:
\begin{equation} \label{Ssch}
\frac{d^2\psi}{dq^2} + \frac{8\pi^2 m}{h^2}\,(E - V)\psi = 0.    
\end{equation}
He then considers the one-dimensional harmonic oscillator: 
\begin{equation}\label{Ham}
H=\frac{P^2}{2m}+\frac{1}{2}m\omega^2 q^2,
\end{equation}
with $m=1$ and $\omega = 2 \pi \nu_0$:
\begin{equation} \label{HO}
E = T + V = \frac{p^2}{2} + 2\pi^2\nu_0^2 q^2.    
\end{equation}
$T$ is the kinetic energy of a one-dimensional particle and $V(q)$ is the potential energy of a harmonic oscillator with frequency $\nu_0$. The $2 \pi^2$ appears because Schrödinger (like Planck) used frequency $\nu$ instead of angular frequency $\omega=2 \pi \nu$.

\noindent Inserting the potential \eqref{HO} into the Schrödinger equation \eqref{Ssch} gives:
\begin{equation} \label{eq22}
\frac{d^2\psi}{dq^2} + 
\frac{8\pi^2}{h^2}\big(E - 2\pi^2\nu_0^2 q^2\big)\psi = 0.   
\end{equation}
To simplify this equation, Schrödinger introduced dimensionless constants $a$ and $b$, and a dimensionless coordinate $x$:
\begin{equation} \label{abx}
a = \frac{8\pi^2 E}{h^2}, \qquad
b = \frac{16\pi^4 \nu_0^2}{h^2}, \qquad
x = q\, b^{1/4}.
\end{equation}
With these substitutions, equation \eqref{eq22} becomes:
\begin{equation} \label{Her}
\frac{d^2\psi}{dx^2} + \left(\frac{a}{\sqrt{b}} - x^2\right)\psi = 0.    
\end{equation}
This is the canonical form of the Hermite differential equation.

\noindent By comparing \eqref{Her} with the standard Hermite equation: 
\begin{equation}
\frac{d^2\psi}{dx^2} + (2n + 1 - x^2)\psi = 0,    
\end{equation}
Schrödinger identified the quantization condition:
\begin{equation} \label{2n1}
\frac{a}{\sqrt{b}} = 2n + 1.    
\end{equation}
Substituting back his definitions of $a$ and $b$ \eqref{abx} yields:
\begin{equation}
\frac{a}{\sqrt{b}} = 
\frac{\tfrac{8\pi^2 E}{h^2}}{\sqrt{\tfrac{16\pi^4\nu_0^2}{h^2}}}
= \frac{2E}{h\nu_0},  
\end{equation}
and setting this equal to $2n + 1$ gives:
\begin{equation} \label{En}
\frac{2E}{h\nu_0} = 2n + 1
\quad\Longrightarrow\quad
E_n = \frac{2n + 1}{2}\,h\nu_0.    
\end{equation}
The general solution of \eqref{Her} (up to normalization) is:
\begin{equation}
\psi_n(x) = e^{-x^2/2}H_n(x),
\end{equation}
where $H_n(x)$ are the Hermite polynomials.  
Substituting back the value of the dimensionless coordinate $x$ \eqref{abx}, Schrödinger obtained:
\begin{equation}
\psi_n(q) = 
\exp\!\left(-\frac{2\pi^2\nu_0 q^2}{h}\right)
H_n\!\left(2\pi q \sqrt{\frac{\nu_0}{h}}\right).    
\end{equation}
Since $\omega=2\pi \nu_0$ and $\hbar = h/2\pi$, \eqref{En} becomes:
\begin{equation}
E_n = \hbar\omega\left(n+\tfrac{1}{2}\right),
\end{equation}
which matches the spectrum from matrix mechanics.
These are (up to normalization) the eigenfunctions of the one-dimensional harmonic oscillator, and the corresponding energy levels \eqref{En} reproduce exactly the quantized spectrum obtained in Heisenberg’s matrix mechanics \cite{Schrod2}. 

In 1925, Heisenberg, Born, and Jordan treated the harmonic oscillator using transition amplitudes $q_{nm}$ between discrete energy levels $E_n$ and $E_m$ \cite{Heisenberg1925, Born-Jordan, Born-Heisen-Jordan}.
From the equations of motion and the \emph{sharpened quantum condition} \eqref{sharpen}, they obtained the relation between energy differences and transition frequencies:
\begin{equation}
h\nu_{nm} = E_n - E_m \qquad  \text{with } \nu_{nm} \text{ the Bohr transition frequencies.}
\end{equation}
Applying this to the harmonic oscillator with Hamiltonian \eqref{Ham} and enforcing the consistency of the commutation relation, Heisenberg, Born, and Jordan derived the discrete energy spectrum:
\begin{equation}
E_n = \left(n + \tfrac{1}{2}\right)h\nu_0, \qquad n = 0, 1, 2, \ldots \qquad \text{where} \qquad \nu_0 = \dfrac{\omega}{2\pi}.
\end{equation}

This was a purely algebraic result, with no wavefunctions, only matrix relations.
When Schrödinger solved the differential equation for the same system \eqref{eq22}, he found that only specific values of $E$ make the wavefunction normalizable \eqref{En}. 
This is precisely the same energy spectrum as that obtained by Heisenberg, Born, and Jordan through matrix mechanics.

In his 1926 paper, “On the Relation between the Quantum Mechanics of Heisenberg, Born, and Jordan, and that of Schrödinger,” Schrödinger showed that if we take any observable $A$ (a matrix in Heisenberg’s sense), we can express the mean value of the quantity $A$ at time $t$ as \cite{Schrod5}:
\begin{equation} \label{over}
\overline{A}(t) = \int \psi^{*}(q,t)\, \hat{A}\, \psi(q,t)\, dq.
\end{equation}
This is the average result of many measurements of $A$ on identically prepared systems.
In Schrödinger’s formulation, each physical system is described by a wavefunction $\psi(q,t)$, where $q$ stands for all spatial coordinates.
An observable quantity (such as position, momentum, or energy) is represented by a Hermitian differential operator $\hat{A}$.

To find how the mean value \eqref{over} changes in time, Schrödinger differentiated $\overline{A}(t)$ with respect to $t$ \cite{Schrod5}:
\begin{equation}
\frac{d}{dt}\,\overline{A}(t)
= \int \left(\frac{\partial \psi^{*}}{\partial t}\right) \hat{A}\, \psi\, dq
+ \int \psi^{*} \hat{A} \left(\frac{\partial \psi}{\partial t}\right) dq
+ \int \psi^{*} \frac{\partial \hat{A}}{\partial t}\, \psi\, dq.
\end{equation}
He then substituted the time-dependent Schrödinger equation:
\begin{equation} \label{wave}
i\hbar\,\frac{\partial \psi}{\partial t} = \hat{H}\psi, 
\qquad
-\,i\hbar\,\frac{\partial \psi^{*}}{\partial t} = \hat{H}\psi^{*},
\end{equation}
and obtained:
\begin{equation} 
\frac{d}{dt}\,\overline{A}(t)
= \frac{i}{\hbar} \int \psi^{*}\, (\hat{H}\hat{A} - \hat{A}\hat{H})\, \psi\, dq
+ \int \psi^{*}\, \frac{\partial \hat{A}}{\partial t}\, \psi\, dq.
\end{equation}
This simplifies to \cite{Schrod5}:
\begin{equation} \label{Schrod1926}
\frac{d}{dt}\,\overline{A}(t)
= \frac{i}{\hbar} \int \psi^{*}\, [\hat{H}, \hat{A}]\, \psi\, dq
+ \int \psi^{*}\, \frac{\partial \hat{A}}{\partial t}\, \psi\, dq.
\end{equation}
This is the same dynamical law that Heisenberg had obtained algebraically for matrices:
\begin{equation} \label{Heis}
\frac{dA}{dt} = \frac{i}{\hbar}[H, A] + \frac{\partial A}{\partial t}.
\end{equation}

Thus, Schrödinger showed that the differential form of his wave mechanics leads to the same operator relations 
as Heisenberg’s matrix equations. The mean values of quantities evolve in time in precisely the same way 
as Heisenberg’s matrix elements. The two theories differ only in their representation. One theory uses differential operators acting on continuous functions, while the other employs infinite matrices acting on discrete sequences.

\subsubsection{Ehrenfest}

In 1926, Paul Ehrenfest derived the relation between the time derivative of expectation values and the commutator of the corresponding operators, using Schrödinger’s differential equation. Ehrenfest’s notation is purely Schrödingerian. 
In his paper, he writes \cite{Ehrenfest}:
\begin{equation} \label{QP}
Q(t) = \int_{-\infty}^{+\infty} x\, \Psi(x,t)\Psi^{*}(x,t)\,dx, 
\qquad
P(t) = i\hbar \int_{-\infty}^{+\infty} \Psi(x,t)
\frac{\partial \Psi^{*}(x,t)}{\partial x}\,dx,
\end{equation}
which are the expectation values of position and momentum, although he did not use that term—it was not yet a standard term.

He differentiates these expressions with respect to time $t$, substituting the time-dependent Schrödinger equation and its complex conjugate:
\begin{align} \label{dift}
i\hbar\,\frac{\partial \Psi}{\partial t} &= 
\left[-\frac{\hbar^{2}}{2m}\frac{\partial^{2}}{\partial x^{2}} 
+ V(x)\right]\Psi, \\
-\,i\hbar\,\frac{\partial \Psi^{*}}{\partial t} &= 
\left[-\frac{\hbar^{2}}{2m}\frac{\partial^{2}}{\partial x^{2}} 
+ V(x)\right]\Psi^{*}.
\end{align}
Using partial integration and assuming that the wavefunction vanishes at infinity (so boundary terms drop out), differentiating again, and using the potential term, Ehrenfest showed:
\begin{equation} \label{45}
\frac{dQ}{dt} = \frac{1}{m}P, \qquad 
m\,\frac{d^{2}Q}{dt^{2}} = 
\int_{-\infty}^{+\infty} \Psi^{*}(x,t)
\left(-\frac{\partial V}{\partial x}\right)\Psi(x,t)\,dx.
\end{equation}
Equation \eqref{45} is the \emph{Ehrenfest theorem}. It shows that the expectation value of the force equals the mass times the acceleration of the expectation value of position.
Ehrenfest’s goal was to show how classical equations of motion emerge as averages from quantum mechanics.
Ehrenfest explicitly remarks that this procedure leaves 
“nothing left out,” emphasizing that the result follows 
directly from Schrödinger’s equation without approximation \cite{Ehrenfest}.

In Dirac’s bra–ket notation, the same relations given in \eqref{45} can be expressed as:
\begin{equation} \label{Ehrhar}
\frac{d}{dt}\langle x\rangle = \frac{\langle p\rangle}{m},
\qquad
m \frac{d^2 \langle x \rangle}{dt^2} = \frac{d}{dt}\langle p\rangle = \left\langle -\frac{\partial V}{\partial x} \right\rangle.
\end{equation}
Here, the angle brackets $\langle\,\rangle$ denote the \emph{expectation value} of an operator in a given quantum state; for any observable $A$, $\langle \psi | \hat{A} | \psi \rangle$, where $| \psi \rangle$ is the state vector of the system. Alternatively, by employing commutators, one obtains:
\begin{equation}
\frac{d}{dt}\langle A \rangle
= \frac{i}{\hbar}\,\langle [H, A] \rangle
+\left\langle \frac{\partial A}{\partial t} \right\rangle.
\end{equation}
This expression is formally identical to \emph{Heisenberg’s equation of motion} \eqref{Heis}.
It states that the time evolution of the expectation value of any observable $A$ is governed by its commutator with the Hamiltonian $H$, together with any explicit time dependence of $A$ itself.

For $V(x) = \frac{1}{2}m\omega^{2}x^{2}$, we have:
\begin{equation}
\left\langle -\frac{\partial V}{\partial x} \right\rangle 
= -m\omega^{2}\langle x \rangle.    
\end{equation}
Hence, Ehrenfest’s results \eqref{Ehrhar} become the Ehrenfest equations for the harmonic oscillator:
\begin{equation}
\frac{d}{dt}\langle x \rangle = \frac{\langle p \rangle}{m},
\qquad
\frac{d}{dt}\langle p \rangle = -\,m\omega^{2}\langle x \rangle.    
\end{equation}
This shows that the expectation values $\langle x \rangle$ and $\langle p \rangle$ obey the classical equations of motion.

Ehrenfest wrote his equations as integrals. Yet they represent the very principle that Dirac expressed in the bra–ket formalism.

\subsubsection{Dirac}

In Section 6 “Bra and Ket Vectors” of Chapter I, “The Principle of Superposition,” in his book \emph{The Principles of Quantum Mechanics}, Dirac presents the bra and ket formalism \cite{Dirac-4}. 
Dirac’s chapter contains the first complete presentation of the bra–ket formalism in print.
Before this, in a 1927 paper, he used bracket-like notations $\alpha'|\alpha''$ for scalar products \cite{Dirac-3}, but had not yet defined “bra” and “ket” as dual vector entities. Hence, Dirac’s 1930 presentation \cite{Dirac-4} is the first to present the symbolic calculus $|\psi\rangle$, $\langle\phi|$, and $\langle\phi|\psi\rangle$ in a mature and self-contained form.

Dirac defines kets as the new type of vector corresponding to the state of a system $|A\rangle$.
Each physical state corresponds to a \emph{direction} of a ket; multiplication by a nonzero complex number
does not change the physical state. The superposition principle is represented by linearity:
\begin{equation}
c_1|A\rangle + c_2|B\rangle = |R\rangle,    
\end{equation}
exhibiting the identification of superposition with vector addition.

Dirac introduces the dual space \cite{Dirac-4}: for every linear functional $\varphi(|A\rangle)$ on kets there exists a bra vector $\langle B|$ such that $\varphi(|A\rangle)=\langle B|A\rangle$. The object $\langle B|$ acting on $|A\rangle$ yields a complex number, the scalar product.
The fundamental linearity and antilinearity relations are \cite{Dirac-4}:
\begin{align}
\langle B|\big(|A\rangle + |A'\rangle\big) &= \langle B|A\rangle + \langle B|A'\rangle,\\
\langle B|\big(c\,|A\rangle\big) &= c\,\langle B|A\rangle,
\end{align}
and addition/scalar multiplication for bras are defined so that \cite{Dirac-4}:
\begin{align}
\big(\langle B| + \langle B'|\big)|A\rangle &= \langle B|A\rangle + \langle B'|A\rangle,\\
\big(c\,\langle B|\big)|A\rangle &= c^{*}\,\langle B|A\rangle.
\end{align}
Dirac also imposes the conjugation relation and positivity:
\begin{equation}
\langle B|A\rangle = \big(\langle A|B\rangle\big)^{*}, \qquad
\langle A|A\rangle>0 \ \ \text{for } |A\rangle\neq 0.    
\end{equation}
These properties define what is essentially a complex inner-product space—the physicist’s version of a Hilbert space—though Dirac’s 1930 presentation is symbolic rather than axiomatic, developed independently of von Neumann’s concurrent mathematical formulation.

Orthogonality and normalization are given by:
\begin{equation}
\langle A|B\rangle = 0, \qquad \langle A|A\rangle = 1,    
\end{equation}
and multiplication by a phase \(e^{i\gamma}\) leaves the physical state unchanged. Dirac explicitly notes that the zero ket
corresponds to no physical state, a key departure from classical analogies (e.g.\ a rest state of oscillation) \cite{Dirac-4}.

Dirac here makes the decisive move from \emph{vectors as arrays of numbers} (matrix mechanics) and
\emph{wavefunctions as functions of coordinates} (wave mechanics) to \emph{state vectors in an abstract vector space} with a dual structure and inner product. This symbolism unifies Schrödinger’s and Heisenberg’s formulations. 

\subsubsection{Stone-von Neumann}

The harmonic oscillator demonstrates that the two epistemologies (matrix and wave mechanics) converge in their outcomes. 
Formally, this unity is guaranteed by the \emph{Stone--von Neumann theorem}. The theorem is completely general in its mathematical statement and not restricted to the harmonic oscillator. While the canonical commutation relations (CCRs) that underlie the theorem appear in the context of the harmonic oscillator, the theorem itself abstracts away from that special case. 

For systems with finitely many degrees of freedom, any regular and irreducible realization of the basic quantum relations between position and momentum, the so-called CCRs, is mathematically equivalent to the standard Schrödinger representation on $L^2(\mathbb{R}^n)$. 
In other words, the abstract algebra of quantum mechanics admits, up to a unitary change of coordinates, only one faithful representation. 
The distinction between Heisenberg’s matrices and Schrödinger’s waves is therefore one of form, not of substance, because both describe the same structure in different symbolic languages.

The result was first outlined by Marshall Stone in 1930, who showed how one can associate a continuous family of unitary transformations with a single self-adjoint operator, thereby giving precise meaning to expressions such as $e^{i t H}$ and laying the groundwork for the modern theory of operators in Hilbert space \cite{Stone1930}. 
Von Neumann, in his 1932 paper, "Über Einen Satz Von Herrn M. H. Stone," generalized and completed Stone’s approach, proving that all consistent representations of the quantum commutation relations are \emph{unitarily equivalent}. 
This theorem formalizes the reconciliation of Heisenberg’s and Schrödinger’s formalisms of the oscillator. In this sense, they are not rival ontologies but two \emph{realizations} of the same algebraic order. 
Both formulations generate identical measurable quantities:
\begin{equation}
\langle \psi(t)|A|\psi(t)\rangle = \langle \psi|A_H(t)|\psi\rangle.
\end{equation}
This equality is a compact Dirac-era bra–ket restatement of what, in earlier formalism, connects Schrödinger’s and Heisenberg’s pictures, and it is conceptually close to Ehrenfest’s theorem, i.e., the application of this relation:
\begin{equation} \label{AH}
\frac{d}{dt}\langle A\rangle
=\frac{d}{dt}\langle \psi(t)|A|\psi(t)\rangle
=\frac{i}{\hbar}\,\langle[H,A]\rangle,
\end{equation}
to position and momentum operators. Von Neumann expressed the equivalence not symbolically but operator-theoretically, in terms of Hilbert-space transformations and unitary conjugations.
Von Neumann formulated this equivalence as a theorem of unitary transformations on a Hilbert space $\mathcal{H}$. 
Let $U_t$ be the unitary group of time evolution generated by a self-adjoint Hamiltonian $H$ on $\mathcal{H}$.
Then for any observable $A$ (a self-adjoint operator on~$\mathcal{H}$) and for all real values of $t$ \cite{vonNeumann1932}:
\begin{equation} \label{NeumanU}
(\psi_t, A \psi_t) = (\psi_0, U_t^{-1} A U_t \psi_0),
\end{equation}
showing that the Schrödinger and Heisenberg pictures are mathematically equivalent and yield identical measurable quantities. 

\noindent In equation \eqref{NeumanU}, $\psi_0$ is the state vector at the initial time $t=0$. 

\noindent The state evolved to time $t$ according to Schrödinger’s equation:
\begin{equation}
i\hbar\,\frac{d}{dt}\psi_t = H \psi_t,    
\end{equation}
is represented by:
\begin{equation} \label{evolve}
\psi_t = U_t \psi_0.    
\end{equation}
$A$ in \eqref{NeumanU} is any observable represented by a self-adjoint operator on $\mathcal{H}$. 

\noindent The unitary time-evolution operator that propagates states forward by $t$ is:
\begin{equation}
U_t = e^{-\tfrac{i}{\hbar}Ht},    
\end{equation}
and the same observable expressed in the Heisenberg picture at time $t$ is:
\begin{equation} \label{AHUT}
A_H(t) = U_t^{-1} A U_t.    
\end{equation}
The inner products in equation \eqref{NeumanU}, $(\psi_t, A \psi_t)$ and $(\psi_0, A_H(t)\psi_0)$, are both taken in the Hilbert-space inner product of $\mathcal{H}$ and represent the expectation value of $A$ at time~$t$, computed in two equivalent ways.
Von Neumann’s formalism thus identifies quantum dynamics with a one-parameter unitary group on $\mathcal{H}$. 
The state vector evolves according to \eqref{evolve} in the Schrödinger picture, or equivalently, the observable evolves according to \eqref{AHUT} in the Heisenberg picture.
Both formulations yield the same physical predictions, differing only in which element of the Hilbert-space triple $(\mathcal{H}, U_t, A)$ carries the time dependence.

Wave mechanics (operators on $L^2$) and matrix mechanics (their components in an orthonormal basis) are related by a unitary change of coordinates; the CCRs and the Stone–von Neumann theorem guarantee equivalence of representations for finite degrees of freedom \cite{vonNeumann1932, Stone1930}.
The Schrödinger and Heisenberg pictures are dynamically unitarily equivalent.

\subsubsection{Born}

Having presented Dirac’s abstract bra–ket formalism and the Stone-von Neumann theorem, we may now turn to Born’s probabilistic interpretation. Born first endowed Schrödinger’s wave function with statistical meaning, while Dirac later cast that insight into the rigorous algebra of bras and kets.

In his 1926 paper, “On the quantum mechanics of collision processes,” Born introduced the probabilistic interpretation of the wavefunction $|\psi|^2$ \cite{Born1, Born2}. Before this paper, both Heisenberg’s matrix mechanics and Schrödinger’s wave mechanics described only stationary states, i.e., energy levels, spectra, and transitions between them. But they did not explain what happens during a process like a collision (e.g., an electron scattering off an atom).
Born’s goal was to determine whether Schrödinger’s wave mechanics could account for quantum jumps—transitions between discrete states—purely within the wave framework, without introducing new postulates.

He examined scattering (collision) processes, such as an electron or $\alpha$-particle approaching an atom.  
Before and after the collision, the two systems behave independently of each other. The atom is in a stationary state $\psi_{na}$, and an incident plane wave represents the incoming particle:
\begin{equation}
e^{\,2\pi i(\alpha x+\beta y+\gamma z)} \equiv e^{\,2\pi i\,\mathbf{k}\cdot\mathbf{r}},
\end{equation}
where $\mathbf{k} = (\alpha, \beta, \gamma)$ is the wave vector of the incident plane wave.
During the interaction, the combined system is coupled by a potential $V(x,y,z;q_k)$ and must be described by a single Schrödinger wavefunction $\Psi(\mathbf r, q_k)$ satisfying the stationary wave equation used in \emph{collision theory} (scattering processes):
\begin{equation} \label{Schroding}
\left(-\frac{h^2}{8\pi^2\mu}\nabla^2 + V\right)\Psi = E\Psi,
\end{equation}
where $\mu$ is the reduced mass of the two-body system.
The full three-dimensional Schrödinger equation describes the \emph{combined system} consisting of the colliding electron (or $\alpha$-particle) and the atom. $\Psi(\mathbf{r}, q_k)$ depends on both the spatial coordinates of the incoming 
particle $\mathbf{r} = (x, y, z)$ and the internal coordinates of the atom $q_k$. This equation describes how the \emph{total wavefunction} of the entire system evolves, 
including both the projectile and the atom’s internal states.
Far from the interaction region, $\Psi$ must reduce to the incident plane wave.

Born constructs the solution of \eqref{Schroding},
showing that after the collision, the total wavefunction becomes a superposition of an incoming wave and outgoing components associated with different atomic final states:
\begin{equation} \label{BornSchematic}
\Psi(\mathbf r,q_k) \sim \text{(incoming wave)} +
\sum_m \Phi_{nm}(\hat{\mathbf r})\,\psi_m(q_k).
\end{equation}
Each complex coefficient $\Phi_{nm}$ is a \emph{scattering amplitude} that measures how strongly the atom in initial state $n$ is scattered into a final state $m$. At the same time, the projectile is deflected in direction $\hat{\mathbf r}$. The amplitude represents both the magnitude and the phase of the outgoing component.

\noindent We compare Born’s exact asymptotic scattering form:
\begin{equation} \label{BornExact}
\Psi(\mathbf r,q_k)
\sim
e^{2\pi i\,\mathbf{k}\cdot\mathbf r}
+
\sum_{m}
f_{nm}(\hat{\mathbf r})\,\frac{e^{ikr}}{r}\,\psi_m(q_k),
\end{equation}
with the schematic expression \eqref{BornSchematic}.  
The two expressions are equivalent once we identify the incoming wave in \eqref{BornSchematic} with the incident plane wave of \eqref{BornExact}:
\begin{equation}
\text{(incoming wave)} \equiv e^{2\pi i\,\mathbf{k}\cdot\mathbf r},
\end{equation}
and define the schematic (compressed) amplitude $\Phi_{nm}(\hat{\mathbf r})$ 
as the outgoing spherical-wave component:
\begin{equation} \label{PhiRelation}
\Phi_{nm}(\hat{\mathbf r})
\equiv
f_{nm}(\hat{\mathbf r})\,\frac{e^{ikr}}{r},
\end{equation}
where $\dfrac{e^{ikr}}{r}$ is the standard outgoing spherical-wave factor 
in the asymptotic form of the scattering solution.

Born recognized that the scattering amplitudes cannot be interpreted causally.  
Quantum mechanics does not predict the specific outcome of an individual collision but only the \emph{likelihood} of possible outcomes.  
He therefore proposed that the probability for the transition $n\!\rightarrow\! m$ into a given solid angle is proportional to the \emph{modulus squared} of the amplitude \cite{Born1,Born2}:
\begin{equation}
P_{n\to m} \propto |\Phi_{nm}|^2.
\end{equation}
This is the first clear statement of the \emph{Born rule}, the fundamental principle that the square of the wave amplitude gives the probability density of an observable result. In Born’s view, quantum mechanics provides a complete but inherently statistical description of atomic processes: deterministic wave evolution yields probabilistic measurement outcomes.

\subsubsection{Pauli}

From Born’s statistical interpretation, the next step was taken by Wolfgang Pauli, who introduced the concept of spin and sought to demonstrate the underlying unity between the matrix and wave formalisms of quantum mechanics.
Pauli’s 1927 paper, “Zur Quantenmechanik des magnetischen Elektrons,” introduces spin operators and endeavors to demonstrate the equivalence of the two formulations: The Heisenberg–Born–Jordan algebraic matrix mechanics and Schrödinger’s wave mechanics.

In Section \S2 of his paper “On the quantum mechanics of the magnetic electron,” Pauli defines spin operators:
\begin{equation} \label{P2}
s_x, s_y, s_z,    
\end{equation}
acting on a two-component wavefunction $(\psi_\alpha, \psi_\beta)$.
He writes matrices (up to conventions and units), and notes that these satisfy the same algebra as the Heisenberg–Jordan matrices.
In Sections \S4–\S5, Pauli applies the operators to concrete Hamiltonians (spin in a magnetic field, the hydrogen atom with spin–orbit and Zeeman terms, and
many-electron symmetry), and demonstrates that this operator–matrix treatment is equivalent to that of Heisenberg and Jordan.
Pauli’s 1927 formulation thus remains fully within the Schrödinger eigenfunction method—wavefunctions and differential operators \cite{Pauli}:
\begin{equation} \label{P1}
p_k = -i\hbar\, \dfrac{\partial}{\partial q_k},    
\end{equation}
but augments it with the operator variables \eqref{P2} acting on a two-component spinor wavefunction \cite{Pauli}:
\begin{equation} \label{P3}
\Psi = (\psi_\alpha, \psi_\beta)^{T},
\end{equation}
and the spin operators \eqref{P2} act as:
\begin{equation}
s_i = \frac{\hbar}{2}\,\sigma_i \qquad (i=x,y,z),
\end{equation}
with the $\sigma_i$ obeying $[\sigma_i,\sigma_j]=2i\,\varepsilon_{ijk}\sigma_k$.

Pauli used Schrödinger’s eigenfunction method, i.e., he wrote wavefunctions $\psi_\alpha(\mathbf{q}), \psi_\beta(\mathbf{q})$ and differential operators for momenta \eqref{P1}. However, he introduced $2 \times 2$ matrix operators \eqref{P2} acting on the two-component spinor \eqref{P3}. 
These operators satisfy the exact algebra as the Heisenberg–Jordan matrices (noncommuting quantities obeying specific commutation relations). So what Pauli showed is that if you extend Schrödinger’s method to include such matrix operators, then the results are mathematically equivalent to those of Heisenberg and Jordan’s matrix mechanics. However, he implemented this equivalence entirely within the framework of the Schrödinger differential equation. He was bridging the two worlds, not abandoning one for the other.

One may call it a synthesis. The Pauli equation is Schrödinger’s method generalized to matrix-valued wavefunctions, and Pauli explicitly says it yields the same physical consequences as the matrix formalism \cite{Pauli}.

\subsection{Hilbert Space}

\subsubsection{Von Neumann}

Although the demonstrations by Schrödinger, Pauli, and others showed striking agreement between matrix and wave mechanics in specific solvable systems—such as the harmonic oscillator and the hydrogen atom—this did not amount to a \emph{general proof of equivalence}. As Fred Muller has demonstrated in his historical–philosophical analysis, the 1926–1927 arguments were not mathematically conclusive, nor could they have been. The two formulations remained only partially aligned, differing in mathematical structure, empirical scope, and ontological commitments. A fully rigorous equivalence emerged only later, with John von Neumann’s 1932 Hilbert-space formulation, which unified them within a single state–observable framework. Until then, the “equivalence” remained, in Muller’s words, “a historically contingent myth” \cite{Muller1997}.

The search for a truly unified and mathematically rigorous foundation did not end with Schrödinger and Dirac. It reached its culmination in the work of von Neumann, whose contributions between 1927 and 1932 transformed the heuristic operator calculus of early quantum mechanics into the abstract and complete structure of Hilbert space. Dirac’s \emph{Principles of Quantum Mechanics} (1930) contains no reference to von Neumann, even though von Neumann had already published his 1927 paper and several follow-up studies. The omission reflects a difference in method and purpose: Dirac developed a symbolic calculus of states and observables based on physical intuition, whereas von Neumann built a rigorous mathematical theory grounded in measure theory and functional analysis.

Von Neumann unified the contributions of Born, Heisenberg, Jordan, Schrödinger, and Dirac into one coherent theoretical structure. He showed that matrix and wave mechanics are not different theories but different \emph{representations} of the same abstract Hilbert-space framework. Schrödinger’s differential operators and Heisenberg’s infinite matrices become coordinate descriptions of self-adjoint operators acting on that space. Von Neumann’s 1927–1932 papers and his 1932 book constitute the first mathematically rigorous unification of all existing forms of quantum mechanics.

In his 1927 paper, “Mathematische Begründung der Quantenmechanik,” von Neumann begins by observing explicitly that matrix mechanics and Schrödinger’s differential method both address the same eigenvalue problem in different languages. The matrix method employs \emph{unendliche Matrizen} (infinite matrices), while wave mechanics uses functions $\psi(q)$ on a domain $\Omega$. He then constructs a common space containing both \cite{Neumann1927}:
\begin{equation} \label{infinit}
\sum_{i=1}^{\infty} |x_i|^2 < \infty
\qquad\text{and}\qquad
\int_{\Omega} |\psi(q)|^2\,dq < \infty,
\end{equation}
showing that both are instances of one mathematical entity, \emph{Der Hilbertsche Raum}, the Hilbert space. 
This move was conceptually novel, transforming a patchwork of infinite-dimensional symbols into a precise linear space with an inner product, in which orthogonality, completeness, and convergence are rigorously defined.

Von Neumann next introduced \emph{self-adjointness} (\emph{Selbstadjungiertheit}) as the mathematically correct criterion for an operator to represent an observable. By specifying proper domains and boundary conditions, he ensured that self-adjoint operators have real spectra and generate unitary time evolution \cite{Neumann1927}. This replaced the physicists’ informal use of Hermitian operators with a rigorous concept that became the cornerstone of functional analysis in quantum theory.

He then defined \emph{unitär} (unitary) transformations as those preserving the inner product \cite{Neumann1927}:
\begin{equation}
\sum_{\mu} a_{\mu\nu} a_{\mu\rho}^{*} = \delta_{\nu\rho}.
\end{equation}
This is the first mathematically precise definition of a unitary operator, formalizing Dirac’s and Jordan’s transformation theory \cite{Born-Jordan} without relying on $\delta$-function heuristics or ambiguous probability amplitudes.

Von Neumann also criticized Dirac’s use of improper eigenfunctions and $\delta(x - x')$. He showed that such objects are not legitimate functions within Hilbert space and argued instead for the use of spectral measures—the idea he later developed into the spectral theorem in his 1932 book \emph{Mathematische Grundlagen der Quantenmechanik} \cite{Neumann1927, Neumann1932}. The novelty in 1927 was not the final spectral theorem but the recognition that the continuous spectrum can be handled rigorously without $\delta$-functions.

He rephrased both matrix and differential eigenvalue equations as the single operator equation $Hf = a f$, where $f$ belongs to a Hilbert space $\Phi$, consisting of sequences or square-integrable functions satisfying \eqref{infinit} \cite{Neumann1927}.
This abstraction marks the birth of the spectral viewpoint: all observables are linear operators on a Hilbert space.

Physicists such as Born, Jordan, and Dirac had the correct physical ideas—probabilities, amplitudes, unitary equivalence—but their mathematics was loose. Von Neumann’s work provided the necessary rigor: proving convergence, defining operator domains, and formalizing the scalar products and norms that underlie the theory.

In his 1932 book, von Neumann gave the first fully rigorous unification of wave and matrix mechanics. He began from the Riesz–Fischer isomorphism between the sequence space of matrix mechanics \cite{Neumann1932}:
\begin{equation}
F_Z = \left\{\, x = (x_\nu) \;\big|\; \sum_{\nu=1}^{\infty} |x_\nu|^2 < \infty \,\right\},
\end{equation}
and the wave-mechanical space:
\begin{equation}
F_\Omega = L^2(\Omega)
= \left\{\, \phi : \Omega \to \mathbb{C} \;\big|\; \int_\Omega |\phi(q)|^2\, dq < \infty \,\right\}.
\end{equation}
which are linearly isometric:
\begin{equation}
\sum_{\nu} |x_\nu|^2 = \int_\Omega |\phi(q)|^2\,dq.
\end{equation}
Under this correspondence, the basic observables match. In the wave picture, position and momentum act by:

\begin{equation}
q_j, \qquad \frac{\hbar}{i}\,\frac{\partial}{\partial q_j},
\end{equation}
while in the sequence space, they correspond to matrices $Q_j, P_j$. Both satisfy the canonical commutation relations \cite{Neumann1932}:
\begin{equation}
[Q_m,Q_n]=0,\qquad [P_m,P_n]=0,\qquad [P_m,Q_n]=i\hbar\,\delta_{mn}\,I,
\end{equation}
and any Hamiltonian $H$ formed algebraically from them yields identical predictions in either representation.

Von Neumann then abstracts to define \emph{Hilbert space} as a complete, separable complex inner-product space. He analyzes orthonormal sets, expansions:
\begin{equation}
f = \sum_\nu (f,\phi_\nu)\,\phi_\nu,
\end{equation}
and completeness. He systematizes operators, their domains, adjoints $A^\ast$, and the distinction between bounded and unbounded operators. The stationary Schrödinger equation becomes the abstract eigenvalue problem $H\phi = \lambda\phi$ for a self-adjoint operator $H$.

He formulated probabilities as quadratic forms \cite{Neumann1932}:
\begin{equation}
\mathbb{P}_\psi\{A \in \Delta\} = (\psi,\,E_A(\Delta)\psi),
\end{equation}
where $E_A(\Delta)$ is the spectral projection of $A$.
For the position operator $Q$ on $L^2(\mathbb{R}^3)$, this reduces to Born’s rule:
\begin{equation}
\mathbb{P}_\psi\{Q \in \Delta\} = \int_\Delta |\psi(x)|^2\, dx.
\end{equation}
Thus the abstract formula $(\psi,E_A(\Delta)\psi)$ gives a representation-independent foundation for Born’s probability law.

In this way, von Neumann completed the transition from quantum mechanics as a calculational scheme to an axiomatized physical theory. States become Hilbert-space vectors (or density operators); observables become self-adjoint operators; probabilities arise from spectral measures. Matrix mechanics and wave mechanics appear as two \emph{unitarily equivalent} representations of a single Hilbert-space theory.

Unlike Dirac \cite{Dirac-1, Dirac-2}, von Neumann does not use bra–ket notation; instead, he writes inner products as $(f,g)$ and works with operators and spectral measures on Hilbert space.

\subsubsection{Schrödinger} \label{Schrodinger1}

Schrödinger criticized both von Neumann’s Hilbert-space formulation and Heisenberg's uncertainty relations (see section \ref{HER}). His 13 July 1935 letter to Einstein \cite{Meyenn}—written in the same months in which he was composing "Die gegenwärtige Situation in der Quantenmechanik" (the "cat paper") \cite{cat}—shows that these objections form a single, coherent interpretive stance. The letter is the private, polemical counterpart to the more polished philosophical exposition in the cat paper. Taken together, they articulate Schrödinger’s consistent rejection of identifying the $\psi$-function with physical reality and his conviction that the standard formalism retains an unexamined classical residue in the way it connects the operator calculus to empirical quantities.

The primary target is von Neumann’s Hilbert-space unification and its associated idealizations. In the letter to Einstein, Schrödinger explicitly directs his critique at von Neumann’s \emph{Mathematische Grundlagen der Quantenmechanik} \cite{Neumann1932}. Although he describes von Neumann as "bei weitem der klarste und sauberste von allen Quantenmechanikern" ("by far the clearest and most rigorous of all quantum mechanicians"), he argues that this clarity is achieved at the cost of importing classical assumptions about the quantities that are supposed to be read off in an experiment \cite{Meyenn}. 

Schrödinger’s central complaint is that von Neumann continues to treat measurement outcomes exactly as if they were values of classical \emph{Bestimmungsstücke}—defining components of a classical model—possessing sharp, determinate boundaries, even though quantum mechanics otherwise abandons the classical picture of particles with definite positions, momenta, and energies. Several examples make this clear.

Von Neumann represents an inexact position measurement by the spectral projector:
\begin{equation}
E([a,b]) \;=\; \text{projection onto the position interval } [a,b].
\end{equation}
This construction assumes that the interval $[a,b]$ is sharply defined, that the particle either \emph{is} or \emph{is not} in that exact interval, and that the measurement outcome corresponds to this classical geometric region of space. Schrödinger objects that no real detector separates space into perfectly sharp intervals: the boundaries of all actual position measurements are, as he puts it, "\emph{verwaschene Ränder}" (blurred edges). Von Neumann’s formalism thus treats the outcome as if it revealed a classical spatial attribute of the particle—precisely the kind of \emph{Bestimmungsstück} Schrödinger rejects.

In von Neumann’s framework, a measurement of the Hamiltonian $H$ yields the value $E_{n}$ with probability $|c_n|^{2}$, and this is modeled by the sharp projection:
\begin{equation}
P_n \;=\; \text{projector onto the exact eigenspace } E_n.
\end{equation}
Here again, the formalism assumes that the system possesses a classical-like energy piece $E_n$, that the spectrum forms sharp, exact bins, and that the apparatus reveals which exact bin is realized. Schrödinger notes that real spectroscopic measurements never produce exact eigenvalues but only finite-width spectral lines; the ideal of perfect spectral resolution is a physical fiction. Von Neumann nevertheless treats the energy readout as sharply defined—another classical \emph{Bestimmungsstück}.

Momentum measurements are represented in exact analogy with position:
\begin{equation}
E([p_1,p_2]) \;=\; \text{projection onto the sharp momentum interval } [p_1,p_2].
\end{equation}
But as Schrödinger emphasizes, real momentum determinations rely on finite apertures, spatial separations, and detectors with limited and fuzzy spectral response. The sharp spectral projector is therefore an idealized classical fiction: it represents the momentum as if it occupied a sharply bounded interval in phase space, even though no actual apparatus achieves such precision.

Schrödinger refers to von Neumann’s treatment of inexact measurements in the closing pages of the \emph{Mathematische Grundlagen} \cite{Neumann1932}. There, von Neumann assumes that the spectrum of an observable can be partitioned into sharply defined intervals and that an experimental reading picks out one such interval with complete definiteness. Only in a footnote does he concede that "tatsächlich sind die Ränder verwaschen," in reality the boundaries are blurred \cite{Meyenn}. For Schrödinger, this is not a minor approximation but a symptom of a deeper issue: the Hilbert-space formalism unifies matrix and wave mechanics by treating ideal sharp spectral decompositions as if they directly correspond to what can be empirically observed, while acknowledging that real situations never achieve such sharpness. The operator calculus may be mathematically impeccable, but the idealizations mediating its empirical interpretation are precisely what Schrödinger finds conceptually suspect \cite{Meyenn}.

This critique reappears in the cat paper in a more expository form. There, Schrödinger stresses that quantum theory still presents itself as if the relevant quantities were classical variables—position, momentum, and energy—attached to a classical model of the system, even though the formalism itself rests on discontinuities and noncommutativity. Yet, he insists, the $\psi$-function does not describe a physical state of the world; it functions instead as an \emph{Erwartungskatalog}, a catalogue of expectations for classical-style outcomes \cite{cat}. "Die Psi-Funktion wird als Erwartungskatalog bezeichnet und erklärt" ("The $\psi$-function is designated and explained as a catalogue of expectations"), he writes to Hans Berliner on 25 July 1935 \cite{Meyenn}. On this view, a superposition is not a physical coexistence of states but a compact description of conditional probabilities for future empirical readings. Schrödinger thereby demotes the ontological significance of \emph{superposition}: it belongs to the expectation-catalog, not to the ontology of the system.
The letters to Einstein and Berliner, together with the cat paper, thus articulate a unified interpretive stance.

The Einstein letter also shows how Schrödinger wove Heisenberg’s uncertainty relations into this critique. He begins by characterizing the new theory as built around "Unbestimmtheitsbehauptung" and "Akausalitätsbehauptung"—claims of indeterminacy and acausality. Yet, he observes that measurement readouts are still represented as definite values (or sharply bounded intervals) on a scale. For Schrödinger, the tension is structural: if the dynamics abandon classical determinacy, the formalism should not tacitly retain classical ideals of sharp numerical outcomes at the empirical interface \cite{Meyenn}.

Schrödinger then asks whether a quantity so idealized-but-finite can meaningfully be paired with a canonically conjugate partner satisfying the exact commutation relations \eqref{sharpen}–\eqref{sharpen2}, which underwrite the usual derivation of the uncertainty relations. He emphasizes that every real device has only finitely many resolvable outcomes; real measurement operators therefore correspond to finite-dimensional approximations of observables, not to perfectly sharp self-adjoint operators on an infinite-dimensional Hilbert space. Schrödinger calls these finitely many possibilities "verabredet" (stipulated in advance).

He then invokes a fundamental mathematical fact: for finite matrices, the trace of a commutator always vanishes, $\mathrm{Tr}(AB-BA)=0$. If the canonical relation $[p,q]=i\hbar I$ held in finite dimensions, then taking the trace of both sides would yield:
\begin{equation}
\mathrm{Tr}(pq-qp)=0
\quad \text{but} \quad
\mathrm{Tr}(i\hbar I_{N}) = i\hbar N,    
\end{equation}
which is impossible for any finite $N$ \cite{Meyenn}. The canonical commutation relation, therefore, cannot hold for any finite-dimensional measurement model. In Schrödinger’s view, the uncertainty relations thus describe the structure of an idealized operator calculus, not the coarse-grained quantities accessed by real apparatuses with finite resolution.

In the same spirit, Schrödinger remarks that he does not know what a $\gamma$-microscope (Heisenberg’s illustrative device) really measures, but that it certainly does not measure a variable possessing a strictly canonical conjugate in the strong operator sense. The uncertainty relations, therefore, cannot be read as transparent ontological claims about reality being intrinsically fuzzy; they express instead the predictive limitations inherent in the $\psi$-function and the operator algebra—constraints on the expectation-catalog, not on the underlying "Merkmale" of the system \cite{Meyenn}.

A further revealing passage in Schrödinger’s 13 July 1935 letter to Einstein shows that he had already grasped the essential structure of what he would soon call \emph{Verschränkung} (entanglement).%
\footnote{The Einstein–Podolsky–Rosen argument (EPR), although historically intertwined with Schrödinger’s response and the origin of the term \emph{Verschränkung}, lies outside the scope of the present paper.} 
In one of the opening paragraphs, Schrödinger introduces a deliberately playful example involving two correlated systems, which he labels the “American” and “European” systems. Schrödinger remarks that his Copenhagen interlocutors repeatedly attempted to reassure him that the correlations involved no "Magie" (magic) in the dramatic sense. Their point was that the distant ("American") Subsystem does not yield $q=6$ or $q=5$, depending on which observable is measured on the proximal ("European") subsystem. No local measurement choice, they insisted, can be used to manipulate or control the outcome statistics of the remote system.
Schrödinger replies that such extreme behavior is not required for the situation to be conceptually problematic. Even if no information can be transmitted, the correlations still allow one to steer the distant subsystem into different pure states by choosing different measurements on the near subsystem---into a $q$--sharp state, or into a $p$--sharp state, or into some other state not belonging to that class. For Schrödinger, this ability to alter the state assignment of the remote system purely through local interventions remains a form of "Magie," even if it does not constitute actionable signaling. As he puts it in his letter to Einstein \cite{Meyenn}:%
\footnote{"Immer wieder wiesen mir die anderen nach, daß keine Magie im krassen Sinne vorliegt ...; immer wieder sagte ich: so arg braucht es nicht zu sein, um blöd zu sein. Ich kann eben doch durch Malträtieren des europäischen Systems das amerikanische willkürlich in einen '$q$-scharfen' Zustand oder in irgendeinen, der bestimmt nicht zu dieser Klasse gehört, hineinsteuern, z.B. in einen '$p$-scharfen.' Das ist auch Magie!"}

\begin{quote}
Again and again the others demonstrated to me that no magic exists in the crude sense ...; and again and again I replied: it does not have to be that extreme to be absurd. By maltreating the European system, I can indeed steer the American one at will into a "$q$-sharp" state, or into some other state that certainly does not belong to that class—for example, into a "$p$-sharp" one. That is magic too!
\end{quote}

Schrödinger describes the phenomenon of entanglement, supplemented by quantum steering. 
The essential point is that the distant system’s post-measurement state depends on the \emph{choice} of measurement performed on its partner, even though no signalling or causal influence is transmitted. Schrödinger emphasizes that this alone is “Magie”—a conceptual tension he finds intolerable. Even without superluminal communication, the ability to steer a remote system into incompatible sharp states ($q$-sharp vs.\ $p$-sharp) simply by choosing what to measure locally already reveals, for him, the fundamentally nonclassical character of the correlations at hand.
This passage is therefore the first clear articulation of the phenomenon Schrödinger would soon baptize \emph{Verschränkung} in the published “cat paper” later in 1935 and the associated \emph{steering}, in the paper he submitted a month later \cite{Schrod1935}. 

The expectation-catalog interpretation is developed even more explicitly in the cat paper \cite{cat}. There, Schrödinger repeatedly insists that the $\psi$-function is neither a physical wave in space nor a complete description of an individual system. Instead, it is a compact device for describing the probabilities of various classical-type measurement results. On this view, the uncertainty relations constrain only the internal structure of this probabilistic catalogue, specifying how the predicted distributions of noncommuting quantities (such as position and momentum) are interrelated. They do \emph{not} assert that these quantities themselves are indeterminate or valueless in reality. Instead, they express the limits of what the expectation-catalog allows one to forecast about the outcomes of classical-style measurements.

In the Einstein letter, Schrödinger reinforces this point by stressing that discrete spectra—such as quantized energies or angular momenta—must reflect real, underlying features of the physical system. In his words, they indicate “neue Merkmale des Systems … in deren Wesen es begründet ist, daß sie nur diese Werte haben können” (“new characteristics of the system … whose very nature requires that only these values are possible”) \cite{Meyenn}. Schrödinger’s point is that these discrete values cannot merely be artifacts of the operator calculus; they must be grounded in structural properties of the system itself. But the Hilbert-space formalism, with its abstract operators and eigenvalue equations, does not reveal what those deeper determining features are. Instead, it supplies a powerful symbolic machinery that predicts the probabilities with which these discrete values will appear, without ever identifying the physical features that make those values necessary in the first place.

Taken together, the letter and the cat paper reveal that Schrödinger’s dissatisfaction with the orthodox formalism is fundamentally epistemic. The $\psi$-function does not, for him, describe what there is; it functions as an expectation-catalog for classical-style quantities. The canonical commutation relations and their uncertainty constraints characterize a highly idealized operator calculus on an infinite-dimensional Hilbert space. In contrast, real experimental arrangements are finite, coarse-grained, and structurally incapable of realizing exact canonical pairs. The surplus of representational content contained in the superposed $\psi$-description—the "neue Merkmale des Systems"—therefore outruns what the operator machinery can recover.

In the language developed later in the paper, one could say that von Neumann transformed quantum mechanics into a beautifully \emph{algorithmic} procedure on Hilbert space—but only by incorporating idealized measurement primitives into the algorithm that Schrödinger regards as physically fictitious. The laws of motion become procedural outputs of a symbolic system, and Schrödinger’s letter amounts to the claim that these procedures rely on unrealistic assumptions about what measurements can, in fact, determine. The operator algebra gives us a robust prediction algorithm, but its connection to actual measurement devices—finite, coarse-grained, and noisy—is precisely where Schrödinger locates the conceptual flaw.

In this sense, Schrödinger already formulates, in 1935, the epistemic asymmetry that will later be recast in algorithmic and complexity-theoretic terms in the subsequent sections: the formalism of quantum mechanics is a powerful symbolic algorithm for predicting structured patterns of outcomes, but it does not exhaust the underlying reality whose features it only partially represents.

In his reply of 8 August to Schrödinger, Einstein tells him, with warmth and genuine sympathy, that he alone is capable of seeing both “from inside and from outside,” yet insists—with characteristic stubbornness—that the $\psi$–function can never describe an individual system but only an ensemble. In the end, he gently steers Schrödinger back to solid ground: whatever the formal elegance of the expectation-catalog may be, no amount of interpretive finesse can turn a superposition of “exploded” and “not-exploded” gunpowder into a real physical state \cite{Meyenn}.%
\footnote{In his reply of 8~August~1935, Einstein introduces a macroscopic reductio of the $\psi$--description that closely anticipates the logic of Schrödinger’s cat. He considers a pile of gunpowder in a chemically unstable equilibrium. Under the unitary dynamics, he observes, the $\psi$--function necessarily evolves into a superposition of “not yet exploded’’ and “already exploded"---a mixture that cannot possibly 
represent the real macroscopic state of a single physical system. For Einstein, this indicates that the $\psi$-function cannot be a complete description of the world, but must be understood statistically, as describing an ensemble of systems rather than an individual one. The example is not a reference to Schrödinger’s later cat paradox; it is 
Einstein’s own macroscopic analogue of the EPR argument, and it likely provided one of the conceptual seeds from which Schrödinger developed the cat illustration in the \emph{Naturwissenschaften} papers later that year \cite{cat}.}

\section{Algorithmic Quantum Mechanics} \label{Alg}

\subsection{The Harmonic Oscillator Example} \label{Har}

Within the framework of matrix mechanics, knowledge is derived from the application of operational rules.
There is no hidden picture behind the computation; performing the operations constitutes knowing.
To understand the state of a system is to grasp how its observables evolve under the Heisenberg equation of motion \eqref{Heis}.

Heisenberg made this epistemic shift explicit in his 1925 paper.
Rejecting unobservable classical quantities, he wrote that earlier quantum theories “contain, as basic element, relationships between quantities that are apparently unobservable in principle, e.g., position and period of revolution of the electron. Thus, these rules lack an evident physical foundation” \cite{Heisenberg1925}.
In his new formulation, “the path of the electron” disappeared; only observable quantities—those appearing in measurable transitions such as frequencies and intensities—remained.
This marked the conceptual pivot from representation to computation.

Matrix mechanics can thus be seen as an algorithmic epistemology—a framework in which operational procedures replace imagery.
The system’s state evolves through deterministic, rule-governed transformations of matrices, and the observer’s task is to carry out these operations rather than to visualize trajectories.
Knowledge, in this sense, is obtained through the execution of the theory’s rules themselves.

Heisenberg’s algorithmic construction is evident in the simple example of the harmonic oscillator. Consider the 
Hamiltonian \eqref{Ham} $(Q \equiv q)$. We compute:
\begin{equation}
[H, Q]
=\frac{1}{2m}[P^2,Q]+\frac{1}{2}m\omega^2[Q^2,Q]
=\frac{1}{2m}\big(P[P,Q]+[P,Q]P\big) 
=-\frac{i\hbar}{m}P.
\end{equation}
Hence $\dot Q=\dfrac{i}{\hbar}[H,Q]=\dfrac{P}{m}$. Assuming no explicit time dependence of $Q, P$:
\begin{equation} \label{H-A}
\frac{dA}{dt}=\frac{i}{\hbar}[H,A].
\end{equation}
Then:
\begin{equation}
[H,P]=\frac{1}{2m}[P^2,P]+\frac{1}{2}m\omega^2[Q^2,P]
=\frac{1}{2}m\omega^2\big(Q[Q,P]+[Q,P]Q\big)
=i\hbar\,m\omega^2 Q,
\end{equation}
so:
\begin{equation}
\frac{dP}{dt}=\frac{i}{\hbar}[H,P]=-\,m\omega^2 Q,
\end{equation}
and:
\begin{equation}\label{QPM}
\dot Q=\frac{P}{m},\qquad \dot P=-\,m\omega^2 Q.
\end{equation}

We have thus derived the quantum analogue of the classical relation $\dot{q} = \dfrac{p}{m}$ using only algebraic manipulation—a purely rule-following process.
No appeal was made to a physical picture of a mass on a spring, to motion in space, or to any wave representation.
The law of motion \eqref{QPM} follows entirely from the formal structure of commutators and algebraic rules.

This demonstration shows why matrix mechanics can be regarded as a form of algorithmic physics.
Physical law emerges from a finite sequence of symbolic operations: applying commutator identities, simplifying expressions, and obtaining the equations of motion.
States are represented by matrices, observables by operators, and dynamics by the execution of algebraic transformations.
The theory dispenses with continuous imagery; its results are generated through procedure rather than depicted through representation.

In matrix mechanics, the law is obtained by carrying out these formal operations.
A finite chain of commutator manipulations yields the dynamical relations, and meaning arises through the process itself.
Classical mechanics depicts what exists—a particle occupying a definite position—whereas matrix mechanics constructs what can be known by calculation.
Understanding, in this framework, consists in applying the transformation rules that govern the operators.

The same operational logic governs Schrödinger’s wave mechanics once reformulated in operator form.
Expressed in Dirac’s algebraic notation, Schrödinger’s picture becomes another realization of algorithmic physics.
To make the equivalence explicit, we may again consider the harmonic oscillator.

In the Schrödinger picture, the derivation of motion remains procedural, but it is revealed through the evolution of expectation values.
It proceeds via Ehrenfest’s theorem, which connects the two pictures by differentiating expectation values under the Schrödinger equation.
With time-independent $Q, P$ and:
\begin{equation} \label{schrodinger}
i\hbar\,\frac{\partial}{\partial t}\,|\psi(t)\rangle=H\,|\psi(t)\rangle,
\end{equation}
Ehrenfest’s theorem yields, for any $A$ without explicit $t$ \eqref{AH}. With $A=Q,P$:
\begin{equation}\label{QPM-Sch}
\frac{d}{dt}\langle Q\rangle=\frac{1}{m}\langle P\rangle,\qquad
\frac{d}{dt}\langle P\rangle=-\,m\omega^2\,\langle Q\rangle,
\end{equation}
identical in content to \eqref{QPM}. 

When the Schrödinger picture is expressed in operator form, it reproduces the same rule-governed dynamics that define Heisenberg’s framework.
The commutator algebra remains the generative structure; only the interpretive standpoint changes—from the transformation of operators to the evolution of states and expectation values.
Formally, both pictures follow identical computational rules, yet epistemically they differ in how knowledge is accessed.
Heisenberg’s formulation produces understanding through procedural transformation, whereas Schrödinger’s, in its original wave form, presented a representational image of motion in space and time.
Recast in Dirac’s algebraic notation, Schrödinger’s formulation relinquishes its representational aspect and assumes the same operational character.
The wave description $\psi(x,t)$ is replaced by the state vector $|\psi(t)\rangle$, whose time evolution obeys the unitary law:
\begin{equation}
|\psi(t)\rangle = e^{-\,iHt/\hbar}\,|\psi(0)\rangle.
\end{equation}
Their unity is algebraic; their difference lies in epistemic orientation—between picturing a process and performing it.

The harmonic oscillator thus offers a minimal yet revealing example of algorithmic physics: a framework in which physical laws arise from the execution of formal operations rather than from pictorial representation.
Heisenberg’s matrix mechanics exemplifies this approach as a rule-based epistemology in which knowledge is obtained through algebraic transformation.
Schrödinger’s original wave mechanics, by contrast, conveyed a continuous visual image of the system’s evolution; only through its reformulation in Dirac’s bra–ket notation does it attain the same procedural form.
In both cases, the essence of quantum theory resides not in visual depiction but in symbolic operation—the law is known through the process that enacts it.

The historical irony is that quantum mechanics achieved formal unification (through Dirac and von Neumann) long before the philosophical asymmetry it embodied was resolved.
The equivalence of the two formalisms is mathematical, not necessarily ontological.
What remains unsettled is whether they describe the same reality or merely compute the same outcomes.
Matrix mechanics eschews spatial ontology, treating observables algebraically; wave mechanics offers a spatially extended representation.
Their difference lies in epistemic style rather than in agreed metaphysics.
Thus, while they are formally unified—like two isomorphic computational models—their epistemological commitments remain distinct, akin to the difference between syntax and semantics in logic.
Mathematics solved what philosophy had not yet fully formulated.
The formal equivalence established by von Neumann and Dirac resolved the mathematical discrepancies between the two formalisms but left their epistemic asymmetry untouched.
Despite formal unification, an epistemic asymmetry remains: Heisenberg emphasizes operator transformation rules; Schrödinger emphasizes a representational state description.

\subsection{Hilbert-von Neumann-Nordheim: The Algorithm of Wave Mechanics} \label{HvNN}

The procedural character of quantum theory is nowhere more evident than in the 1928 paper by David Hilbert, John von Neumann, and Lothar Nordheim, "Über die Grundlagen der Quantenmechanik" \cite{Hilbert}.
Their formulation reconstructs Schrödinger’s differential equations as the direct outcome of a finite sequence of symbolic operations.
It is an early instance of \emph{algorithmic physics}: a rule-based derivation of the laws of motion from axioms about operators, without pictorial representation or reference to trajectories.

In modern terms, what Hilbert, von Neumann, and Nordheim achieved was the conversion of wave mechanics into a deterministic computational algorithm.
The Schrödinger equation appears not as an assumption but as a theorem obtained through mechanical manipulation of operator relations:

\begin{enumerate}

\item \textbf{Canonical basic operators, defined in section \S 5 (Axiom)} \cite{Hilbert}:
\begin{equation} \label{32}
p\,f(x)=\varepsilon\,\frac{\partial f}{\partial x}, \qquad
q\,f(x)=x\,f(x), \qquad
\varepsilon=\frac{h}{2\pi i} \quad (\varepsilon = i \hbar).    
\end{equation}
Equation \eqref{32} establishes the algebraic form of the fundamental observables—the quantum analogues of the classical
variables $p$ and $q$.

\item \textbf{Canonical conjugacy (Axiom)}:
\begin{equation} \label{before 64}
GF-FG=1.    
\end{equation}
This commutation (canonical) relation is the structural postulate underlying all of quantum kinematics.

\item \textbf{Amplitude equations for Hermitian $F$, defined in section \S 6 (Axiom—fundamental postulate)} \cite{Hilbert}:
\begin{equation} \label{64a}
(F^{(x)}-y)\,\varphi(xy;\,qF)=0,
\end{equation}
\begin{equation} \label{64b}
(G^{(y)}+\varepsilon\,\partial_y)\,\varphi(xy;\,qF)=0.
\end{equation}
Here $G$ is canonically conjugate to $F$.  
These equations constitute the defining dynamical rule for the amplitudes $\varphi(xy;\,qF)$; from them, the Schrödinger equations are derived.

\medskip
\emph{At this point, the formal scaffolding of the theory is complete.  
Items (1)–(3) establish an abstract rule-system: algebraic operators, their canonical relation, and the amplitude postulate that will generate dynamics.  
The following steps illustrate how concrete physical quantities—energy and time—fit into this purely symbolic framework.}

\item \textbf{Energy and time (Definition).}
Let $F=H(p,q)$ be the energy (Hamiltonian) and $t$ its canonically conjugate variable.
This defines the energy observable and its conjugate, identifying the physical quantities to which the axioms are applied.

\medskip
\emph{Having introduced the physical observables, the next step is deductive: applying the abstract axioms to these variables yields the familiar equations of quantum mechanics.  
In what follows, the Schrödinger equations will appear not as postulates but as consequences of the formal operator calculus.}

\item \textbf{Derived law $\Rightarrow$ Stationary Schrödinger equation.}
We choose $F=H(p,q)$ (energy), insert this into \eqref{64a}, and set $y=W$ (the energy value). This yields \cite{Hilbert}:
\begin{equation} \label{70}
\Biggl\{\, H\, \!\Bigl(\varepsilon\,\frac{\partial}{\partial x},\, x\Bigr) - W \,\Biggr\}\, \varphi (xW;\, qH) = 0.
\end{equation}
\emph{Here the formal machinery produces the stationary Schrödinger equation automatically, showing that what is usually taken as a foundational assumption is, in this framework, a derived theorem.}

This equation represents the stationary Schrödinger equation, derived algorithmically from the
Hilbert--von Neumann--Nordheim amplitude postulate \eqref{64a}.  
It expresses the condition that the amplitude
$\varphi(xW;\,qH)$ is an eigenfunction of the Hamiltonian operator
$H\!\big(\varepsilon\tfrac{\partial}{\partial x},\,x\big)$
with eigenvalue $W$, the possible value of the energy observable $H$.

\noindent For discrete eigenvalues $W=W_n$ \cite{Hilbert}:
\begin{equation} \label{71}
\varphi(xW_n;\,qH)=\psi_n(x),
\end{equation}
which expresses that the functions $\psi_n(x)$ are the stationary eigenfunctions of the Hamiltonian corresponding to the quantized energy levels $W_n$.

\medskip
\emph{Once the stationary case is obtained, the procedural logic extends seamlessly to the temporal domain.  
By introducing the conjugate variable $t$, the same formal algorithm generates the time-dependent law.}

\item \textbf{Derived law $\Rightarrow$ Time-dependent Schrödinger equation.}
Now we use the conjugate variable $t$, apply equation \eqref{64b} with $F=H$ and $y=t$, and define  $\psi(x,t)\:=\varphi(xy;\,qt)$. This yields \cite{Hilbert}:
\begin{equation} \label{72}
\Biggl\{\,
H\, \!\Bigl(\varepsilon\,\frac{\partial}{\partial x},\, x\Bigr)
+\varepsilon\,\frac{\partial}{\partial t}
\Biggr\}\,\psi(x,t) = 0 .
\end{equation}
Hilbert, von Neumann, and Nordheim used a sign convention opposite to the modern one:

\begin{equation} \label{72-1}
\Biggl\{\, H\!\Bigl(\varepsilon\,\frac{\partial}{\partial x},\, x\Bigr)
-\varepsilon\,\frac{\partial}{\partial t}
\Biggr\}\,\psi(x,t)=0,
\qquad \text{which yields} \qquad
i\hbar\,\frac{\partial \psi}{\partial t} = H\psi .
\end{equation}

\medskip
\emph{From these two differential laws—the stationary and time-dependent forms—the complete procedural structure of quantum mechanics emerges.  
The next derivation makes explicit how discrete and continuous spectra arise within this unified formalism.}

\item \textbf{Derived consequence $\Rightarrow$ Spectral form.}
The same procedure yields the discrete and continuous expansions \cite{Hilbert}:
\begin{equation} \label{73}
\psi(x,t)
= \sum_{n} c_{n}\,\psi_{n}(x)\,
e^{\,\tfrac{2\pi i\, W_{n}t}{h}}.
\end{equation}
Immediately afterward, Hilbert, von Neumann, and Nordheim remark that \cite{Hilbert}:
\begin{equation}
\bigl|\,c_{n}\,e^{\,\tfrac{2\pi i\,W_{n}t}{h}}\bigr|^{2}=|c_{n}|^{2},
\end{equation}
interpreting $|c_{n}|^{2}$ as the probability that the system is in the $n$-th stationary state.

Again, Hilbert, von Neumann, and Nordheim used a sign convention opposite to the modern one. The historical sign gives: 

\begin{equation}
\frac{2 \pi i W_n t}{h},   
\end{equation}
instead of the modern expectation:
\begin{equation}
\frac{- i W_n t}{\hbar}.   
\end{equation}

\medskip
\emph{At this stage, the formal calculus reproduces the statistical interpretation of quantum mechanics:  
Probabilities emerge as invariants of the amplitude’s spectral decomposition.}

The most general form of the stationary solution, encompassing both discrete and continuous spectra,
is then given by \cite{Hilbert}:
\begin{equation} \label{74}
\varphi(xW;\,qH)
= \sum_{n} c_{n}\,\psi_{n}(x)\,\delta(W - W_{n})
+ \varphi_{c}(x;W),
\end{equation}
where the first term represents the discrete part of the energy spectrum and the second term $\varphi_{c}(x; W)$ represents the continuous component.
Equation \eqref{74} unifies the discrete and continuous cases within a single formal expression for the amplitude $\varphi(xW;\,qH)$.
It provides the spectral decomposition of this amplitude, combining the discrete spectrum (bound states $W_n$) with the continuous spectrum (scattering states $\varphi_c$).
Mathematically, the expression represents the most general solution of the stationary Schrödinger equation, showing that both bound and continuum energy states arise from the same operator rule derived from the Hilbert–von Neumann–Nordheim axioms.

\medskip
\emph{The sequence thus completes an algorithmic cycle: from axioms to definitions, from differential laws to spectral consequences.  
This structure makes clear that the “laws” of quantum theory are procedural outputs of a symbolic system.}

\end{enumerate}

Equations \eqref{70} and \eqref{72} correspond respectively to the stationary and time-dependent Schrödinger equations,
obtained through direct substitution and differentiation rules applied to the operator algebra.
The derivation requires no spatial model or pictorial analogy; only sequential rule application is necessary: operator replacement, differentiation, and symbolic simplification.

Hilbert, von Neumann, and Nordheim’s formalism thus constitutes an early expression of \emph{algorithmic quantum mechanics}.
As in Heisenberg’s matrix mechanics, knowledge emerges through the execution of well-defined algebraic operations.
The wavefunction itself is an intermediate computational object, and the Schrödinger equation represents the rule governing its transformation.
The theory is therefore procedural: its “laws” are generated by applying symbolic operations rather than by depicting motion in space and time.

Algorithmic quantum mechanics, in this sense, unites Heisenberg’s transformation theory and Schrödinger’s wave mechanics under a common operational logic: the laws of motion are obtained not by describing but by \emph{performing} the transformations.

The axiomatic language developed by Hilbert, von Neumann, and Nordheim in 1928 constitutes the formal articulation of the rule-system implicit in Heisenberg’s matrix mechanics.
Heisenberg’s original formulation had already been procedural—it replaced the descriptive imagery of classical mechanics with an algebra of transformations, a calculus of commutators and matrix equations that produced the laws of motion through explicit manipulation.
Hilbert, von Neumann, and Nordheim elevated this computational grammar into an axiomatic structure.
They translated Heisenberg’s rules into a logically organized system of postulates governing operators and amplitudes.
The mathematics of quantum theory thus acquired a deductive syntax: observables became operators, their relations codified in commutation axioms, and their dynamics generated by differential operator equations such as \eqref{70}–\eqref{74}.

However, this axiomatic language did not unify Heisenberg’s and Schrödinger’s formalisms by blending them.
Instead, it \emph{absorbed} Schrödinger’s wave mechanics into Heisenberg’s operational logic.
Within the new framework, $p$ and $q$ are abstract operators satisfying the canonical commutation relation, and the wavefunction $\psi(x,t)$ becomes merely one possible representation of the general amplitude $\varphi(xy;\, qF)$.
The Schrödinger equation, when derived from the amplitude postulates, no longer describes a physical wave propagating in space but an algebraic rule acting in an abstract function space.
In this sense, Schrödinger’s mechanics was not unified with Heisenberg’s, but rather \emph{Heisenbergized}: the pictorial representation was re-expressed as an operator procedure, with its spatial imagery replaced by rule-governed transformations.

The result is a profound epistemic shift.
The axiomatic formalism generalizes Heisenberg’s procedural epistemology while dissolving Schrödinger’s representational one.
Computation becomes the mode of understanding; algebraic transformation replaces physical depiction.
The unification achieved by Hilbert, von Neumann, and Nordheim is thus not ontological but linguistic: it translates all of quantum mechanics into the single syntactic language of operator algebra, where to know a system is to perform its transformations.

\section{Complexity and the Epistemic Divide} \label{Comp}

\subsection{Epistemic Asymmetry and Computational Analogy}

Having traced how Hilbert, von Neumann, and Nordheim transformed Schrödinger’s wave mechanics into an axiomatic operator framework, and how von Neumann subsequently proved its formal equivalence to Heisenberg’s mechanics, we can now ask why their epistemic difference persists—why the procedural and representational modes of understanding remain asymmetrical despite mathematical unification.

I contend that computational complexity theory, particularly the distinction between $\mathsf{P}$ and $\mathsf{NP}$ (and their quantum analogue, $\mathsf{BQP}$), provides a concrete formulation for this epistemic divide.
Problems in $\mathsf{P}$—or more broadly in $\mathsf{BQP}$—are solvable by efficient, rule-based procedures. This corresponds to Heisenberg’s framework, where knowledge arises through the execution of well-defined operations.
By contrast, $\mathsf{NP}$ comprises problems whose proposed solutions can be efficiently verified once supplied, reflecting Schrödinger’s framework, in which knowledge appears as a recognizable structure rather than as a constructed process.

Throughout this discussion, the $\mathsf{P}$–$\mathsf{NP}$ comparison functions as an epistemic analogy, not a formal reduction. It contrasts procedural construction—algorithms that find—with recognitional verification—procedures that check. No claim is made that wave mechanics itself belongs to any specific complexity class. The mapping Heisenberg $\leftrightarrow$ $\mathsf{BQP}$ and Schrödinger $\leftrightarrow$ $\mathsf{NP}$ serves only as a conceptual heuristic about modes of knowing, not as a mathematical assertion.

Complexity theory, I argue, renders explicit the gap between constructing and recognizing—two forms of cognition that differ in their practical efficiency. What began as a difference in epistemic style has evolved into a measurable limitation grounded in both physics and computation.

\subsection{Computation (Heisenberg), Interference, Outcome (Schrödinger)} \label{HIS}

I argue that quantum computation offers a partial bridge between these epistemologies.
Algorithms in the class $\mathsf{BQP}$ (defined in footnote \ref{class}) exploit interference and superposition to render certain recognitional tasks constructively accessible—effectively implementing Heisenberg-style procedural dynamics that reveal Schrödinger-style patterns.
In a quantum algorithm, the system evolves through a sequence of unitary operations, each step governed by deterministic mathematical rules.

During this evolution, amplitudes corresponding to different computational paths combine through interference.
If the problem possesses an internal structure—such as a periodic or symmetrical pattern—the interference can be engineered so that the correct solution’s amplitude reinforces itself while incorrect ones cancel out.
This in-between stage of interference constitutes the epistemic bridge: the lawful dynamics (Heisenberg) are transformed into a recognizable pattern (Schrödinger).

When the final measurement collapses the state, the correct outcome emerges with high probability.
Thus, recognition itself becomes the product of computation—knowledge that in a classical setting would merely be verifiable is here procedurally realized through the evolution of the system.

The epistemic trajectory of a quantum algorithm can thus be represented schematically as a three-stage process, bridging Heisenberg’s procedural construction and Schrödinger’s recognitional revelation through the mediating role of interference (see table \ref{T1}).
\begin{table}[ht!]
\centering
\caption{Epistemic sequence in quantum computation}
\begin{tabular}{p{3cm}p{5cm}p{4cm}p{4cm}}
\toprule
\textbf{Stage} & \textbf{Physical / Mathematical Action} & \textbf{Epistemic Mode} & \textbf{Analogy} \\
\midrule
\textbf{Heisenberg (Procedural)} & 
The quantum system evolves through a series of unitary transformations — rule-governed, deterministic operations on the state space. &
\emph{Construction:} Knowledge is produced by performing lawful operations. &
Algorithmic execution: step-by-step evolution of the quantum state. \\

\textbf{Interference (Bridge)} &
Amplitudes from many computational paths combine; the structure of the problem determines which paths interfere constructively or destructively. &
\emph{Translation:} The system’s internal dynamics encode recognitional content into the amplitude pattern. &
The dynamic stage where physics "searches" through structure without representing it pictorially. \\

\textbf{Schrödinger (Recognitional)} &
Measurement collapses the wavefunction, revealing a definite outcome. The interference pattern ensures that the correct result appears with high probability. &
\emph{Recognition:} The structure becomes visible---the answer is \emph{seen} or verified. &
The output is an intelligible representation: the outcome reveals what was previously hidden. \\
\bottomrule
\end{tabular} \label{T1}
\end{table}

\newpage
Interference serves as the physical and epistemic bridge: the lawful dynamics of the computation (Heisenberg) reorganize amplitude structure so that, upon measurement (Schrödinger), the correct result becomes recognizably manifest.

\subsection{The White Raven and a Needle in the Haystack}

This epistemic sequence—from procedural construction through interference to recognitional outcome—has a formal counterpart in the hierarchy of computational classes.
In particular, while quantum computation extends the constructive reach of what can be efficiently computed, current evidence suggests that its power remains limited:

\begin{equation} \label{Heir}
\mathsf{P} \subseteq \mathsf{BQP} \subseteq \mathsf{PP} \subseteq \mathsf{PSPACE},
\quad
\mathsf{NP} \not\subseteq \mathsf{BQP} \ \text{(as a belief, not a theorem)}.
\end{equation}

This hierarchy reflects a progressive increase in theoretical computational power.%
\footnote{\label{class} $\mathsf{P}$ (Polynomial Time): classical, deterministic algorithms, what ordinary computers can do efficiently.

\noindent $\mathsf{BQP}$ (Bounded-Error Quantum Polynomial Time): quantum algorithms that can err slightly but can reduce error by repetition. This class explores the computational capabilities enabled by quantum mechanics.

\noindent $\mathsf{PP}$: allows probabilistic computation with arbitrarily small biases, so loose that it becomes vastly more powerful.

\noindent $\mathsf{PSPACE}$: everything we can do if only memory, not time, is bounded.  Time refers to the number of computation steps (such as seconds of processing). Space means the amount of memory or storage (e.g., RAM, tape length) used during computation. It means any problem solvable with polynomial space, even if that takes exponential time. This is extremely powerful because we solve problems by recursively exploring all possibilities one at a time, without having to remember all of them simultaneously; for example, deciding who wins in a game like chess or Go on an $n \times n$ board. We could theoretically search the entire game tree by recursion, using only enough memory to store the current path. A limited amount of space can be recycled indefinitely, so space-bounded algorithms can sometimes simulate computations that would seem to require infinite time.
This makes $\mathsf{PSPACE}$ extremely powerful because it contains $\mathsf{P}$, $\mathsf{NP}$, and $\mathsf{BQP}$.
First, we argue:
\begin{equation} \label{Heir1}
\mathsf{P} \subseteq \mathsf{NP} \subseteq  \mathsf{PP} \subseteq \mathsf{PSPACE},
\end{equation}
If an algorithm runs in polynomial time, it cannot possibly use more than polynomial space. 
Hence, every $\mathsf{P}$ algorithm automatically satisfies the polynomial-space constraint. 
Since $\mathsf{PSPACE}$ has limited memory but unlimited time, and $\mathsf{NP}$ requires simulating exponentially many possibilities, this can be done one at a time, sequentially, reusing the same space at each step.
Each inclusion is known to hold (provably). None of the inclusions is known to be strict, but we have strong evidence that they are.}

Most theorists believe $\mathsf{NP}$ is not contained in $\mathsf{BQP}$. $\mathsf{NP}$ contains problems whose solutions can be verified in polynomial time.
To say $\mathsf{NP}\subseteq\mathsf{BQP}$ would mean that a quantum computer could solve every $\mathsf{NP}$ problem efficiently. No one has proven this impossible, but current evidence and decades of research strongly suggest otherwise. No quantum algorithms for $\mathsf{NP}$-complete problems have been found despite massive effort.
Those are the two main, conceptually distinct arguments for why most theorists believe $\mathsf{NP}\nsubseteq\mathsf{BQP}$:

1. The Oracle Argument (the formal, meta-mathematical reason):

Carl Hempel’s \emph{paradox of the ravens} provides an epistemological analogue to what happens with oracle separations in complexity theory. 
Although Hempel’s formulation has been widely criticized in philosophy of science for its counterintuitive implications about confirmation and evidential relevance, its underlying \emph{asymmetry of evidence} remains instructive. Philosophers generally accept that Hempel’s paradox exposes a limitation of purely formal accounts of confirmation, not a prescription for how science should work. However, in the complexity-theoretic analogy, I am not using Hempel’s paradox literally as a model of scientific confirmation, but metaphorically to illuminate asymmetry. I adopt only one side of Hempel’s logic, the falsification asymmetry. One counterexample invalidates a universal claim, while countless confirmations can never prove it conclusively. Thus, even though the confirmation aspect of Hempel’s paradox is philosophically disputed, its logical asymmetry between confirmation and falsification remains sound. And that is precisely the part the analogy with the oracle exploits.

We can now translate this epistemic asymmetry into the formal language of computational theory. An \emph{oracle world} can be imagined as an alternate universe of computation, where every Turing machine---classical, probabilistic, or quantum---has access to the same magical subroutine $A$. In this universe, we can compare complexity classes such as $\mathsf{NP}^A$ and $\mathsf{BQP}^A$, meaning $\mathsf{NP}$ and $\mathsf{BQP}$ relative to the oracle $A$.
Researchers have constructed oracles $A$ for which:
\begin{equation}
\mathsf{NP}^A \not\subseteq \mathsf{BQP}^A.    
\end{equation}
This is the computational analogue of discovering a \emph{white raven}, a concrete counterexample to the universal statement "for all oracles $A$, $\mathsf{NP}^A \subseteq \mathsf{BQP}^A$." 
Once even one such oracle exists, no \emph{relativizing proof}---that is, a proof that remains valid when any oracle is added---can ever establish $\mathsf{NP}\subseteq\mathsf{BQP}$ in the real, non-oracle world. 
A relativizing argument must hold for \emph{every} possible oracle; since a counter-oracle already exists, the universal claim is falsified.

This result constitutes a \emph{meta-proof barrier}. 
It does not determine what quantum computers physically can or cannot do; rather, it identifies the \emph{limits of our mathematical methods}. 
Oracle separations tell us which categories of reasoning are too weak to settle the inclusion of $\mathsf{NP}$ in $\mathsf{BQP}$.

Although oracle results are artificial, i.e., they do not describe our physical universe, they serve a crucial logical purpose. 
They define the limits of the range of proof techniques available, i.e., they mark a boundary beyond which no relativizing argument can reach. 
Thus, the oracle argument does not solve the $\mathsf{NP}$-versus-$\mathsf{BQP}$ question itself, but it sharply defines where proof attempts must transcend relativization to succeed.

2. The Interference Argument (the physical, mechanistic reason): 

This is an intuitive, physics-based reason grounded in how quantum computation actually works. Quantum computation often gets misdescribed as “trying all possibilities at once.”
This is false in the crucial way that matters for $\mathsf{NP}$ problems. Quantum computers explore a superposition of many computational paths, but measurement collapses the state, resulting in only one outcome.
In other words, a quantum computer evolves a superposition of exponentially many basis states, but the amplitudes are complex numbers that must interfere coherently.
We cannot simply read out all branches; measurement yields one outcome, with probability proportional to the squared magnitude of its amplitude.
So, unless the algorithm arranges constructive interference toward the correct answer and destructive interference toward all wrong ones, we cannot gain an exponential advantage. However, finding that interference pattern is itself as hard as solving the problem. 

Shor’s algorithm arranges constructive interference toward the correct answers and destructive interference toward all wrong ones. But this is only possible because the solution space forms a coherent periodic structure. 
For combinatorial problems (SAT Boolean Satisfiability, 3-Coloring) — the archetypes of the $\mathsf{NP}$-complete class — where no global symmetry is known, no known global structure exists. The solution space is irregular, so amplitudes cannot be aligned coherently across exponentially many wrong possibilities.

We need to know where to interfere, that is, to embed the structure of the solution into the Hamiltonian or unitary evolution. Quantum physics will not automatically find a needle in an exponentially large haystack unless we already represent information about the needle’s location.
We thus reach the combinatorial explosion barrier. We can hold exponentially many amplitudes, but cannot extract exponentially many bits of information from them. Quantum mechanics conserves probability amplitude through interference, not information volume. Hence, quantum computation does not eliminate exponential blowup; it reorganizes it. 

Thus, $\mathsf{NP}\not\subseteq\mathsf{BQP}$ is not a theorem. It is a widely held consensus belief, based on overwhelming circumstantial evidence.

The hierarchy \eqref{Heir} shows that computational possibility does not align neatly with logical reducibility.
The existence of a physical procedure (a quantum algorithm) does not entail the collapse of logical distinctions between complexity classes.
Physical realizability, therefore, constrains epistemic accessibility. The reason is that even if a solution can be recognized in principle, it may not be constructible within the physical limits of computation.

In black-hole information theory, this computational asymmetry assumes a strikingly physical form.
Daniel Harlow and Patrick Hayden demonstrated in 2013 that decoding Hawking radiation—i.e., recognizing entanglement across an event horizon—would require computational resources that grow exponentially with the black hole’s entropy.
Even if such correlations are, in principle, recognizable, they are inaccessible in practice due to complexity.
They argued that the quantum computation required to distill the requisite entanglement from Hawking radiation would take time exponential in the black hole entropy, rendering the AMPS (firewall) experiment operationally infeasible \cite{HarlowHayden}.
Nature thus seems to enforce a \emph{physical analogue of the $\mathsf{P}\neq\mathsf{NP}$ gap} because certain truths exist and can be verified in principle but cannot be feasibly constructed.

\subsection{\texorpdfstring{$\mathsf{NP}\not\subseteq\mathsf{BQP}$}{NP not subseteq BQP} and Schrödinger and Heisenberg}

The analogy between $\mathsf{BQP}$ and $\mathsf{NP}$, developed in section \ref{HIS}:

\medskip
\noindent \textbf{Heisenberg} $\Rightarrow$ $\mathsf{BQP}$, \textbf{Schrödinger} $\Rightarrow$ $\mathsf{NP}$,
\medskip

\noindent represents the computational reformulation of the early epistemic divide in quantum theory. 
The methodological contrast between \emph{construction} (Heisenberg) and \emph{recognition} (Schrödinger) reappears as the distinction between procedural and verificational modes of knowing, now expressed through the language of computational complexity. 
Heisenberg’s matrix mechanics embodies the logic of rule-following transformation---knowledge generated through formal operations---just as $\mathsf{BQP}$ describes problems solvable by efficient quantum procedures. 
Schrödinger’s wave mechanics, by contrast, parallels $\mathsf{NP}$: it provides an immediately recognizable structure once the correct form is guessed, even if that form cannot be efficiently derived.

The unifying work of von Neumann and Dirac, who demonstrated the mathematical equivalence of the two formalisms, plays the same role here that the structural hierarchy among $\mathsf{P}$, $\mathsf{BQP}$, and $\mathsf{NP}$ plays in computational theory: it establishes a common formal framework within which distinct epistemic modes coexist. 
Von Neumann’s Hilbert-space formalism thus finds its analogue in the modern complexity hierarchy---it unifies without erasing differences:

\medskip
\noindent \textbf{Von Neumann’s formal unification} $\Longleftrightarrow$ \textbf{The structural hierarchy linking} $\mathsf{BQP}, \mathsf{NP}, \text{ and } \mathsf{P}$.

\noindent Both reconcile systems that are mathematically compatible yet epistemically distinct.

\begin{equation} \label{S-sub-H}
\textbf{Schrödinger} \quad \not\subseteq \quad  \textbf{Heisenberg}  
\end{equation}

Equation \eqref{S-sub-H} embodies this epistemic non-containment in its original historical and philosophical setting. 
It recalls the period between 1926 and 1927, when Schrödinger and Heisenberg were locked in a contest over the correct form of quantum mechanics. 
Schrödinger, convinced that physics required continuity and visualizability, sought to ground the new theory in wave phenomena and spatial representation. 
The matrix theorists, by contrast, reacted with sharp skepticism and even hostility. 
They regarded Schrödinger’s smooth, pictorial formalism as a regression to classical imagery and doubted that such an intuitive theory could capture the quantum discontinuities they had labored to formalize. 
Heisenberg judged it "too good to be true" \cite{Beller1999}. 
Their dispute was not about predictive accuracy but about what \emph{knowing} a quantum system meant. 
Although Schrödinger’s and Heisenberg’s theories were formally equivalent, demonstrated by von Neumann, their orientations toward knowledge remained incommensurable: procedural versus representational, becoming versus being. 
The reconciliation achieved within Hilbert space concealed rather than eliminated this philosophical tension.

This same asymmetry reappears today in computational terms. 
$\mathsf{BQP}$ captures what can be constructed through lawful quantum dynamics---efficiently realizable procedures such as interference and unitary evolution---whereas $\mathsf{NP}$ represents what can be recognized once given but not necessarily built by any feasible process. 
The structural relation $\mathsf{NP}\not\subseteq\mathsf{BQP}$ mirrors the older non-containment \eqref{S-sub-H}. 
Both describe an enduring boundary between what can be \emph{computed} and what can only be \emph{represented}.

Schrödinger’s wavefunction contains complete information about the system, yet that information cannot be accessed without collapse. 
Similarly, an $\mathsf{NP}$ solution can be efficiently verified once presented, but its construction may demand exponential resources. 
Heisenberg’s formulation, conversely, yields only what can be produced through lawful dynamical evolution. 
In this sense, $\mathsf{BQP}$ delineates the scope of what is constructively realizable within quantum mechanics.

The methodological divide of 1925---whether to represent or to compute---thus reemerges as a physical boundary in nature itself. 
It separates what the laws of physics can produce efficiently through evolution from what they can merely contain as latent structure, recognizable but not constructible. 
Viewed through the lens of computational complexity, quantum theory embodies this boundary in physical terms: interference enables efficient computation of structured truths. In contrast, the combinatorial explosion blocks efficient access to arbitrary ones. 
The contrast between the procedural and the recognitional, therefore, reappears as a fundamental asymmetry between what is computable and what is verifiable in the universe.

\noindent Epistemically, the representational viewpoint extends further than the procedural one. 
Their formal equivalence, demonstrated by von Neumann, does not erase this distinction. 
The non-containment is epistemic, not formal: not all representable truths are algorithmically attainable. 
The wavefunction describes \emph{being}; the operator formalism describes \emph{becoming}. 
These two forms of access to reality remain distinct in their efficiency and constructive reach.

The implication is a limit on procedural knowledge: not every representable state of nature can be generated by an efficient physical process, even in quantum theory. 
The Heisenberg picture is confined to what can be realized through computational dynamics; the Schrödinger picture encompasses what exists in principle, independent of constructibility. 
In this way, the asymmetry becomes a principle of epistemic physics: the physically constructible constitutes a proper subset of the recognizably real. 
Quantum theory, far from eliminating this duality, preserves it as a structural feature of the universe itself.

\section{Conclusion}

It is striking that the philosophical divide between the primacy of measurement and the continuity of reality has reemerged—transformed—within modern discussions of information scrambling, reversibility, and circuit complexity.
What Heisenberg and Schrödinger once debated as a matter of interpretation is now examined in laboratories through the behavior of noisy quantum processors.

At the center of these recurrences lies a question of form rather than content—one now expressed in the syntax of computational complexity: is $\mathsf{P} = \mathsf{NP}$?
Heisenberg may be viewed as aligned with $\mathsf{P}$: the procedural constructor who obtains results through the faithful execution of algorithmic rules.
Schrödinger, by contrast, aligns with $\mathsf{NP}$: the recognitional verifier who, once supplied with the correct wavefunction, confirms its validity by direct substitution.
In early quantum theory, this distinction emerged as the difference between computing spectral data in matrix mechanics and recognizing stationary states in wave mechanics—two epistemic routes to the same truth, one constructive and the other immediately verifiable.

Historians have often framed their divergence in ontological terms—discrete versus continuous, symbolic versus pictorial—but seldom as a difference in epistemic or computational mode.
Viewed through the lens of modern complexity theory, the contrast acquires a new clarity.
Heisenberg’s formalism was operational and opaque: one computed observables without visualizing the structure that produced them, much like an algorithm in $\mathsf{P}$ that yields results without exposing its internal path.
Schrödinger’s formulation, conversely, made the structure visible once the correct form was posited, akin to $\mathsf{NP}$ problems where verification is immediate once a candidate solution is given.

Seen in this light, the methodological tension that inaugurated quantum mechanics anticipates the central question of theoretical computer science.
The procedural and recognitional epistemologies of Heisenberg and Schrödinger prefigure the modern asymmetry between construction and recognition that defines computational complexity.
What began as a metaphysical dispute over how nature can be known has evolved into a physical limitation on what nature permits to be computed.

The program of complexity theory can therefore be read as the philosophical culmination of that trajectory—an implicit assertion that the universe itself enforces a kind of procedural realism.
Nature, in this perspective, privileges what can be constructed over what can merely be conceived, turning the epistemological debate of the 1920s into a computational principle of the twenty-first century.

\section*{Acknowledgment}

This essay is dedicated to the memory of my doctoral supervisor, \textbf{Mara Beller} (1945–2004), whose insight into the dialogical nature of scientific revolutions continues to shape our understanding of the conceptual foundations of quantum theory. 
Although more than two decades have passed since our many conversations on the history and philosophy of quantum mechanics, her intellectual presence remains vivid in my work. 
This reconstruction of the Heisenberg–Schrödinger divide owes much to her example of rigorous, historically grounded analysis. 
In this International Year of Quantum Mechanics (2025), I remember with gratitude a scholar whose voice deserves wider remembrance—and whose legacy endures in the continuing dialogue between history, philosophy, and physics.

\end{document}